\documentclass[12pt, a4paper]{article}

\usepackage{subfigure}
\usepackage{graphicx}
\usepackage{amsmath}
\usepackage{amssymb}
\numberwithin{equation}{section}

\else
\fi \topmargin -30pt

\begin{document}
\title{Defects and Dressed Boundaries\\ in Complex Sine-Gordon Theory}
\author{P. Bowcock\footnote{peter.bowcock@durham.ac.uk}\;\; and J. M. Umpleby\footnote{j.m.umpleby@durham.ac.uk} \\   Centre for Particle Theory\\ Department of Mathematical Sciences \\ University of Durham \\ Durham, DH1 3LE, U.K.  }
\maketitle

\begin{abstract}
In this paper we constructed integrable defects in complex sine-Gordon theory. Soliton and particle interactions with the defect are analysed. Defects are used to dress Dirichlet boundaries to create a wider class of integrable boundary conditions to ones previously studied. Properties of these new boundaries  including soliton and particle reflections with the boundaries are presented.
\end{abstract}

\newpage

\section{Introduction}

The study of defects in statistical models has a history going back at least to the 60's \cite{FishF:1967}. More recently there has been some interest in the question of whether it is possible to
introduce defects in the context of integrable $1+1$ dimensional field theories in a way which preserves
the integrability of the model.  It was pointed out a number of years ago \cite{Delfino:1994nx} that for theories with a non-trivial S-matrix, factorisation implied that the defect must either
be purely transmitting or purely reflecting. Purely transmitting defects have been 
studied in the sine-Gordon theory using algebraic \cite{Konik:1997gx} and
Lagrangian methods \cite{Bowcock:2005vs}, and for the nonlinear Schr\"odinger equation \cite{Corrigan:2005iq,Caudrelier:2007} and Affine Toda Theories \cite{Bowcock:2004my}. 

In this paper we will use the Lagrangian approach to study integrable defects in the complex sine-Gordon (CSG) model. The CSG model was first introduced independently by  Lund and Regge as a model of relativistic vortices in a superfluid \cite{Lund:1976ze,Lund:1977dt} and by Pohlmeyer in a dimensional reduction of a $O(4)$ non-linear $\sigma$-model\cite{Pohlmeyer:1979ch}. It belongs to the class of homogeneous sine-Gordon theories, which are $G/U(1)$ gauged Wess-Zumino-Witten Models perturbed by a potential. For CSG the group $G=SU(2)$. More recently, the quantum case has been studied by Dorey and Hollowood\cite{Dorey:1994mg} and Maillet and de Vega\cite{deVega:1982sh}. The CSG model has been studied with a boundary\cite{Bowcock:2002vz}. Boundary conditions were found which preserved integrability and soliton solutions were used to analyse  the classical reflection factor. Subsequently in \cite{Bowcock:2006hj} a conjecture for the fully quantum reflection matrix was given. 

The CSG theory differs from the sine-Gordon model in a number of ways. Like the sine-Gordon equation, the CSG equation possesses soliton solutions, but unlike the sine-Gordon model these are non-topological as the CSG model has a unique vacuum. In addition solitons carry a charge, which is a consequence of the theory's  $U(1)$-invariance. We shall find that defects, like solitons, can carry charge in the CSG model. In the sine-Gordon theory, defects can acquire topological charge in the process of soliton scattering, absorption or emission. Indeed, in the latter two cases, this is crucial for 
the conservation of quantities such as energy; the energy of the soliton is transferred to or from the defect, whose energy is a function of its topological charge. In the CSG we shall find that the defect
can still absorb or emit a soliton. In this case, the energy of the defect can change during the process because of the multivalued nature of the Lagrangian defect potential. 

Most recently the CSG model has attracted some attention in the context of magnons in string theory.
The CSG equation is equivalent to the equations of motion of a string moving on an $\mathbb{R}$ x $S^{3}$ subspace of $AdS_{5}$ x $S^{5}$ \cite{Pohlmeyer:1975nb}. This equivalence is used in current work verifying the prediction of the $A$d$S$/CFT correspondence that the spectrum of operator dimensions in planar $\mathcal{N}=4$ SUSY Yang-Mills and the spectrum of a free strings on $A$d$S_{5}$ x $S^{5}$\cite{Chen:2006gea} are the same. Integrable boundaries have already been used in this context \cite{Hofman:2007xp, Ahn:2007bq}.

In the next section, the complex sine-Gordon Model in the bulk will be summarised, both its formulation as a perturbed Wess-Zumino-Witten Model and the more standard description as a complex field in 1+1 dimensional spacetime with $U(1)$ symmetry. Soliton solutions will be presented with their energy, momentum and charge explicitly noted, along with the B\"{a}cklund transformation used to generate these solutions. A new B\"{a}cklund transformation is also given. This dispenses with the need to 
introduce an auxiliary (dual) field, and facilitates the introduction of defects into the theory. Explicit formulae for
the time-delay and phase shift in soliton-soliton and soliton-particle scattering are presented.

In the following section, we will construct the complex sine-Gordon theory with defect. As a check that  classical integrability is maintained after the introduction of the defect, we construct conserved energy, momentum and charge in the presence of the defect. We then analyse the vacuum of the theory, and discuss soliton and particle scattering with the defect.

In the fourth section we will use the integrable defect to dress the CSG boundary. In \cite{Bowcock:2002vz}, integrable boundary conditions for the CSG model were considered under the restriction that they did not involve time derivatives of the field. Dressing the boundary by placing an integrable defect in front of the Dirichlet boundary, first discussed at \cite{Bowcock:2006}, generates new boundary conditions involving time derivatives of the CSG field. The new boundary conditions are characterised by two parameters, and can emit and absorb solitons of different charges. New expressions for conserved charge and energy  in the presence of a dressed boundary are found, as well as the complete formula for the dressed boundary action. We then analyse particle and soliton interactions with the boundary, find boundary bound states and calculate reflection factors.

In the final section, we offer a summary of the work covered and some discussion and thoughts about future directions.

\section{Complex sine-Gordon theory}
\label{sec:one}
The complex sine-Gordon theory is a 1+1 dimensional field theory described by the following Lagrangian

\begin{eqnarray}\label{eq:bulklagrangian}
 \mathcal{L}_{CSG}& =&  \frac{ \partial_{t} u \partial_{t}u^{*} -\partial_{x} u \partial_{x} u^{*} }{1-  \lambda^{2} u u^{*}} - 4\beta uu^{*} \, .
\end{eqnarray}
Here $u$ is a complex field, $\lambda$ is the coupling constant  and $\beta$ the mass parameter. The constant $\lambda$ can be absorbed into $u$ and $u^{*}$ by scaling the field, in which case it appears as an overall factor multiplying the Lagrangian. In this paper we will only deal with the classical theory, so we can consistently set $\lambda$ equal to $1$. Note that the Lagrangian has a global $U(1)$ symmetry, i.e. it remains unchanged under the transformations $u \rightarrow e^{i\alpha}u, \ u^{*} \rightarrow e^{-i\alpha}u^{*}$; this leads to a conserved charge by Noether's theorem. The Lagrangian looks very similar to that of a massive free complex field. Interactions  in the model can be thought of as arising when the denominator $1-  \lambda^{2} u u^{*}$ is expanded in  small fluctuations about the vacuum $u=0$. The complex sine-Gordon equation of motion (and its complex conjugate)
\begin{eqnarray}\label{eq:bulkeom}
 \partial_{tt} u - \partial_{xx} u  + \frac{u^{*} ((\partial_{t} u)^{2}-(\partial_{x}u)^{2})}{1-uu^{*}} + 4\beta u(1-uu^{*}) &=&0
\end{eqnarray}
are derived by varying the action $S = \int \ dt L$ in the usual way.

The connection to the sine-Gordon theory can be explicitly seen by a change of variables $u = \mathrm{sin}\phi\ e^{2i\eta}$. When $\eta$ is taken to be a constant, the equation (\ref{eq:bulkeom}) becomes the familiar sine-Gordon equation of motion,

 \begin{eqnarray}\label{eq:sgeom}
 \phi_{tt} - \phi_{xx} + 4\beta \mathrm{cos}\phi\mathrm{sin}\phi &=& 0 \, .
\end{eqnarray}

One can straightforwardly derive the forms of the conserved energy, momentum and charge to respectively be

\begin{eqnarray}\label{eq:energy}
 E &=& \int dx \ \ \frac{ \partial_{t}u\partial_{t}u^{*}+\partial_{x}u\partial_{x}u^{*}}{1-uu^{*}} + 4\beta uu^{*}\, ,\nonumber\\
 P&=&-\int dx \ \ \ \frac{\partial_{x}u\partial_{t}u^{*} +\partial_{x}u^{*}\partial_{t}u}{1-uu^{*}}\, , \nonumber\\
Q &=& i \ \int dx \ \frac{u\partial_{t}u^{*}-u^{*}\partial_{t}u}{1-uu^{*}}\, .
\end{eqnarray}
Once again the expressions for these conserved charges deviate from those of the free complex field by the denominator factor multiplying the derivative terms. The theory has two sectors corresponding to the sign of $\beta$, in these two sectors the vacuum field configurations are different. When $\beta > 0$, the vacuum is $u=0$ and therefore topologically trivial, while when $\beta < 0$  the energy is minimised when $|u| = 1$, therefore in this regime the vacuum spontaneously breaks the U(1) symmetry of the theory.

As first demonstrated by Bakas\cite{Bakas:1994} the complex sine-Gordon model can be thought of as a $SU(2)/U(1)$ gauged Wess-Zumino-Witten model (WZW) perturbed by a potential with the action

\begin{eqnarray}
S &=& S_{gWZW} + S_{pot}\, .
\end{eqnarray}
$S_{gWZW}$ is the standard gauged WZW action, defined on a 3-dimensional manifold whose boundary is the normal 2-dimensional space, and $S_{pot}$  is a potential term which breaks conformal invariance and gives masses to the fields. In this formulation the two sectors are treated simultaneously, with the connection between the $SU(2)$ group element in the WZW treatment and the CSG field $u$ given by 

\begin{eqnarray}
 g &=& \left( \begin{array}{cc} u&-iv^{*} \\ -iv & u^{*} \end{array} \right) \, ,
\end{eqnarray}
where $v=-\sqrt{1-uu^{*}}e^{-i\theta}$. Here $\theta$ is an auxiliary field defined up to a constant by the following constraint equations
\begin{eqnarray}\label{eq:thetaconstraint}
\partial_{t}\theta &=& -\frac{i}{2}\left(\frac{u\partial_{x}u^{*}-u^{*}\partial_{x}u}{1-uu^{*}} \right)\, ,\nonumber\\
\partial_{x}\theta &=& -\frac{i}{2}\left(\frac{u\partial_{t}u^{*}-u^{*}\partial_{t}u}{1-uu^{*}} \right)\, ,
\end{eqnarray}
which arise as the equations of motion for the gauge field in the gauged WZW action.
The complex field $u$ satisfies the CSG equation with $\beta >0$ whereas the dual field $v$ satisfies the equation with $\beta <0$. It follows that the vacuum has the form
\begin{eqnarray}
 g_{vac}&=& \left( \begin{array}{cc} 0 & ie^{-i\Omega} \\ ie^{i\Omega} & 0 \end{array}  \right) \, .
\end{eqnarray}
The B\"{a}cklund transformation (BT) from the WZW formulation \cite{Park:1995gc}
\begin{eqnarray}
g^{-1}\partial g -f^{-1}\partial f - \delta \sqrt{\beta} \left[ g^{-1}\sigma f , \sigma \right] &=& 0 \, ,\nonumber\\
\bar{\partial}g g^{-1} \sigma - \sigma \bar{\partial} f f^{-1} +  \frac{\sqrt{\beta}}{\delta}\left[gf^{-1} , \sigma\right] &=& 0 \, ,
\end{eqnarray}
can also be re-written in terms of the complex field picture
\begin{eqnarray}
 -\frac{\partial_{t} w -\partial_{x} w }{z^{*}} + \frac{\partial_{t} u- \partial_{x} u}{v^{*}} +2\sqrt{\beta}\delta\left( wv+uz\right)& =& 0 \, , \nonumber \\
\frac{\partial_{t} u + \partial_{x} u}{v} + \frac{\partial_{t} w + \partial_{x} w}{z} -\frac{2\sqrt{\beta}}{\delta}\left( uz^{*}-wv^{*}\right) &=& 0 \, ,
\end{eqnarray}
where $z =-\sqrt{1-ww^{*}}e^{-i\psi}$ is the dual field to $w$. Another auxiliary field $\psi$ is defined by similar constraint equations
\begin{eqnarray}\label{eq:psiconstraint}
\partial_{t}\psi &=& -\frac{i}{2}\left(\frac{w\partial_{x}w^{*}-w^{*}\partial_{x}w}{1-ww^{*}} \right) \, , \nonumber\\
\partial_{x}\psi &=& -\frac{i}{2}\left(\frac{w\partial_{t}w^{*}-w^{*}\partial_{t}w}{1-ww^{*}} \right) \, .
\end{eqnarray}

 Substituting the vacuum field configuration ($u=0,\ v=-e^{i\Omega}$) into the BT allows the one-soliton solutions $w= u_{1-sol},\ z=v_{1-sol}$ to be generated

\begin{eqnarray}
u_{1-sol} &=&\frac{ \mathrm{cos}(a) \mathrm{exp}(2i\sqrt{\beta}\mathrm{sin}(a)(t\ \mathrm{cosh}(\chi)-x\ \mathrm{sinh}(\chi)))}{\mathrm{cosh}(2\sqrt{\beta}\mathrm{cos}(a)(x\ \mathrm{cosh}(\chi)-t\ \mathrm{sinh}(\chi)))}\, , \nonumber\\
 v_{1-sol} &=& -e^{-i\Omega}\left(\mathrm{cos}(a)\mathrm{tanh}(2\sqrt{\beta}\mathrm{cos}(a)(x\ \mathrm{cosh}(\chi)-t\ \mathrm{sinh}(\chi)))-i\mathrm{sin}(a)\right)\, ,
\end{eqnarray}
where $\delta = e^{\chi}$ and the phase in the dual field $z$ is given by
\begin{equation} \psi = -\Omega + \mathrm{arctan}\left(\mathrm{tan}(a)\mathrm{coth}(2\sqrt{\beta}\mathrm{cos}(a)(x\ \mathrm{cosh}(\chi)-t\ \mathrm{sinh}(\chi)))\right) \, .\end{equation}   
These solutions were originally derived by Getmanov \cite{Getmanov:1977hk} and Lund and Regge \cite{Lund:1976ze}.
Whereas earlier work considered both sectors simultaneously, we have discovered that it is possible to consider just the $\beta >0$ sector of the theory. By combining results from the WZW picture and those of Pohlmeyer \cite{Pohlmeyer:1979ch},  the B\"{a}cklund transformation can be put in the form

\begin{eqnarray}\label{eq:DefectCond}
0&=& \frac{u_{t}-u_{x}}{\sqrt{1-uu^{*}}} - \frac{w_{t}-w_{x}}{\sqrt{1-ww^{*}}}e^{i\alpha} +2\sqrt{\beta}\delta\left(w \sqrt{1-uu^{*}} + u\sqrt{1-ww^{*}} e^{i\alpha}\right)\, ,\nonumber\\ 
0&=& \frac{u_{t}+u_{x}}{\sqrt{1-uu^{*}}}e^{i\alpha} + \frac{w_{t}+w_{x}}{\sqrt{1-ww^{*}}} -\frac{2\sqrt{\beta}}{\delta}\left(u \sqrt{1-ww^{*}} - w\sqrt{1-uu^{*}} e^{i\alpha}\right)\, ,
\end{eqnarray}
where $\alpha$ is given by
\begin{equation}\label{eq:alpha}
\alpha =\mathrm{arcsin}\left[\frac{i}{2}\left(\frac{uw^{*}-wu^{*} +2i\mathrm{sin}A}{\sqrt{1-ww^{*}}\sqrt{1-uu^{*}}}      \right) \right] \, .
\end{equation}
These equations are written just in terms of the fields $u$ and $w$, dispensing with the need for $v$ and $z$. The function $\alpha$ gives an explicit expression for $\alpha \equiv \theta - \psi$ in terms of 
$u$, $w$ and a new parameter $A$. The BT now depend on two parameters $\delta$ and $A$. 

\begin{figure}[htb]
\fbox{
  \centering
  \begin{minipage}[c]{0.45\textwidth}
    \centering
 \resizebox{\textwidth}{!}{
   \begin{tabular}{cc}
		
		$0\ <\ a\ <\ \frac{\pi}{2}$& $ Q_{sol} =\ -4a + 2\pi$\\
		&\\
		$\frac{\pi}{2}\ <\ a\ <\ \pi$ &$ Q_{sol} =\  -2\pi+4a$\\ 
		&\\
		$-\frac{\pi}{2}\ <\ a <\ 0 $& $ Q_{sol} =\ -4a - 2\pi $\\
		&\\
		$-\pi\ <\ a\ <\ -\frac{\pi}{2}$&$ Q_{sol} =\ 4a + 2\pi$\\ 
		
	\end{tabular}}
  \end{minipage}
  \begin{minipage}[c]{0.45\textwidth}
    \includegraphics[width=0.8\textwidth,height=0.8\textwidth]{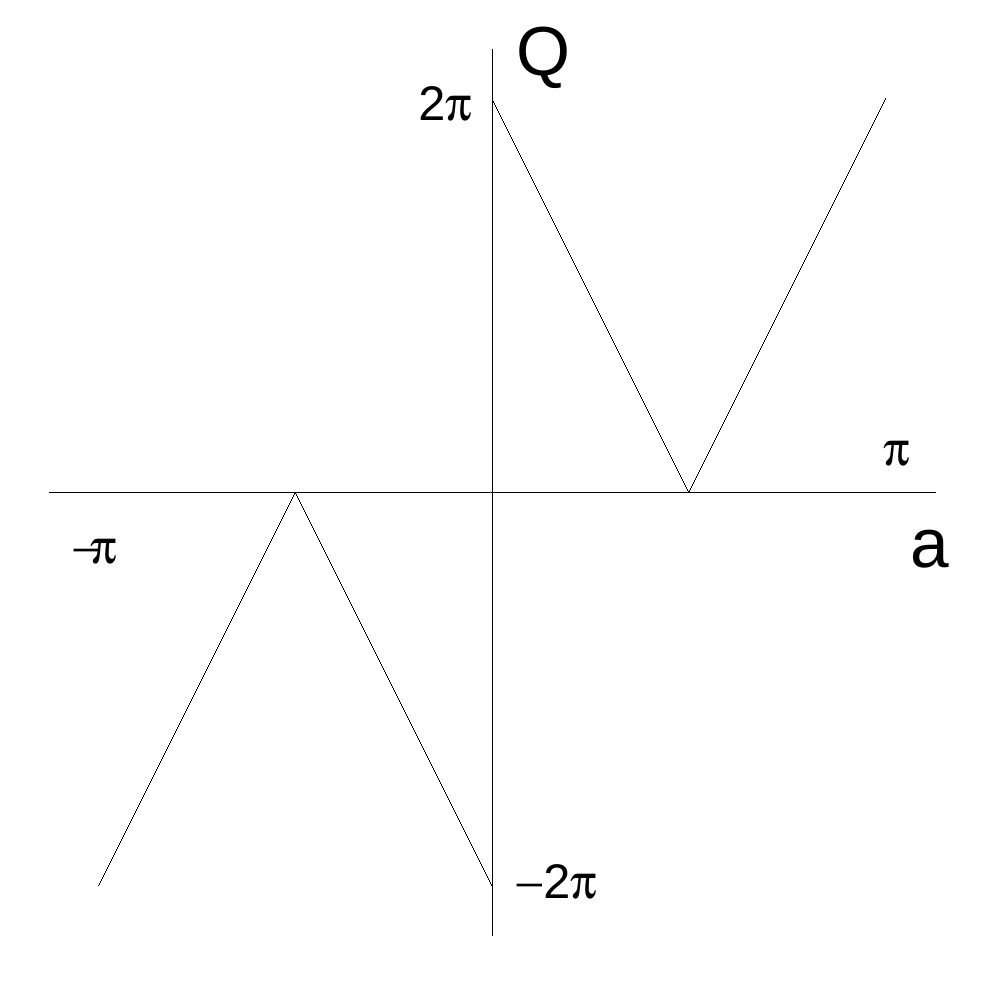}
  \end{minipage}
}
  \caption{Charge $Q(a)$ of the complex sine-Gordon soliton.}\label{fig:solitoncharge}
\end{figure}

As usual, it is straightforward to show that if the complex fields $u$ and $w$ satisfy the BT (\ref{eq:DefectCond}), they automatically satisfy the CSG equation (\ref{eq:bulkeom}). The BT is extremely useful as it simplifies the task of finding soliton solutions to the CSG equation, by reducing the order of the differential equation that needs to be solved. 
The form of a one-soliton solution can be found by substituting $u=0,\ u^{*}=0$ into the BT and solving for $w$, which yields

\begin{eqnarray}
w=u_{1-sol}&=&\frac{ \mathrm{cos}(a) e^{2i\sqrt{\beta}\mathrm{sin}(a)(t\ \mathrm{cosh}(\theta)-x\ \mathrm{sinh}(\theta))}}{\mathrm{cosh}(2\sqrt{\beta}\mathrm{cos}(a)(x\ \mathrm{cosh}(\theta)-t\ \mathrm{sinh}(\theta)))} \, .
\end{eqnarray}
The parameters in the soliton solution are related to the parameters in the BT by $a=A$ and  $ e^{\theta} = \delta$. To solve the BT a choice on the behaviour of $\alpha$, which in this situation has the form \begin{equation}\alpha =\mathrm{arcsin}\left[\frac{-\mathrm{sin}A}{\sqrt{1-ww^{*}}} \right]\, ,  \end{equation} has to be made. We choose that to the left of the soliton solution $\alpha=A \pm \pi$ and to the right $\alpha=-A$. The one-soliton solution has the form of a wave packet with phase velocity  $\mathrm{coth}(\theta)$ and group velocity $\mathrm{tanh}(\theta)$. The factor $\mathrm{cos}(a)$ gives the height of the solution, in the limit $\mathrm{cos}(a) \approx \mathrm{small} $ it reduces to a plane wave solution and in the limit $\mathrm{sin}(a)=0$ the solution becomes real. It is noted that due to the non-topological nature of the vacuum, the soliton is also non-topological. The energy of the soliton is always positive, while the momentum can be positive or negative. They both depend on the charge parameter $a$ and the rapidity of the soliton $\theta$.  

\begin{eqnarray}\label{eq:energysol}
E_{soliton} &=& 8\sqrt{\beta}\left|\mathrm{cos}(a)\right|\mathrm{cosh}(\theta) \nonumber\\
P_{soliton} &=& 8\sqrt{\beta}\left|\mathrm{cos}(a)\right| \mathrm{sinh}(\theta)\ 
\end{eqnarray}
The charge of the soliton $Q_{sol}$, graphically represented in figure \ref{fig:solitoncharge}, is $2\pi$ periodic in $a$.

\begin{figure}[htp!]
\begin{center}
\includegraphics[width=4in]{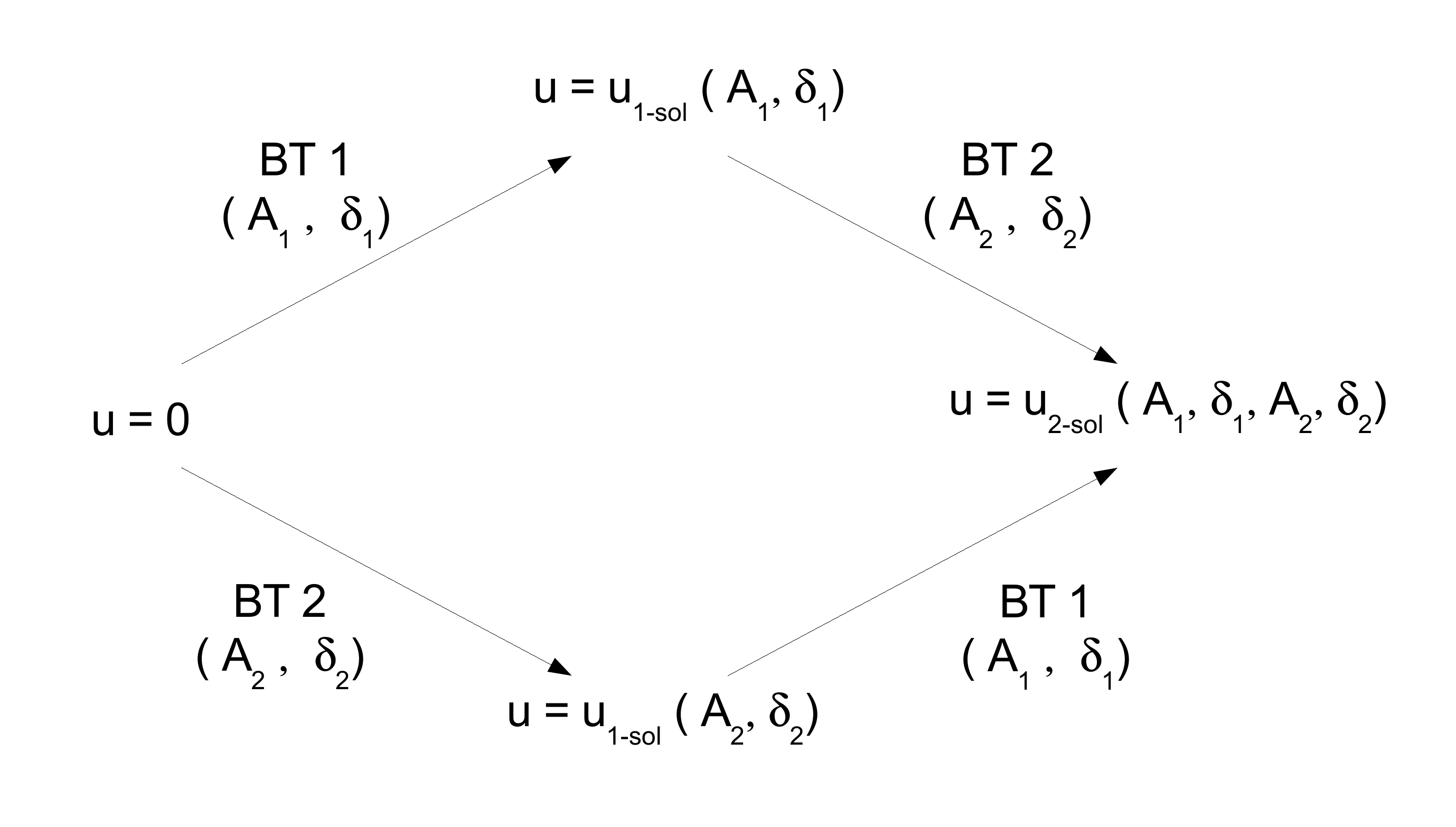}
\end{center}
\label{fig:thmperm}
\caption{Generating multi-soliton solutions via the Theorem of Permutability}
\end{figure}

The BT can be used iteratively to construct multi-soliton solutions. A two-soliton solution can be obtained by taking $u=u_{1-sol}$ and solving for $w$. This procedure can be simplified using the Theorem of Permutability, illustrated in figure \ref{fig:thmperm}, which asserts that the use of multiple BTs with different parameters is a commutative process. This reduces the task to solving purely algebraic equations and gives the two-soliton solution 

\begin{eqnarray}
\label{eq:2sol}
u_{2-sol} &=& \frac{(u_{1}\delta_{1}-u_{2}\delta_{2})(\sqrt{1-u_{2}u^{*}_{2}}e^{-i\alpha_{2}}\delta_{1} -\sqrt{1-u_{1}u^{*}_{1}}e^{-i\alpha_{1}}\delta_{2})}{\delta_{1}^{2}-\delta_{1}\delta_{2}(u_{1}u_{2}^{*}+ u_{2}u_{1}^{*} +2\sqrt{1-u_{1}u_{1}^{*}}\sqrt{1-u_{2}u_{2}^{*}}\mathrm{cos}(\alpha_{1}-\alpha_{2})) + \delta_{2}^{2}}\nonumber\\
&& +\frac{(u_{1}\delta_{2}-u_{2}\delta_{1})(\sqrt{1-u_{1}u^{*}_{1}}e^{i\alpha_{1}}\delta_{1} -\sqrt{1-u_{2}u^{*}_{2}}e^{i\alpha_{2}}\delta_{2})}{\delta_{1}^{2}-\delta_{1}\delta_{2}(u_{1}u_{2}^{*}+ u_{2}u_{1}^{*} +2\sqrt{1-u_{1}u_{1}^{*}}\sqrt{1-u_{2}u_{2}^{*}}\mathrm{cos}(\alpha_{1}-\alpha_{2})) + \delta_{2}^{2}} \, ,\nonumber\\
&&  
\end{eqnarray}
where $u_{1}$ and $u_{2}$ are the two constituent one-solitons
\begin{eqnarray}
\label{eq:1sol}
u_{i} &=& \frac{N_{i}\mathrm{cos}(a_{i}) e^{2i\sqrt{\beta}\mathrm{sin}(a_{i})(\mathrm{cosh}(\theta_{i})t-\mathrm{sinh}(\theta_{i})(x-c_{i}))}}{\mathrm{cosh}(2\sqrt{\beta}\mathrm{cos}(a_{i})(\mathrm{cosh}(\theta_{i})(x-c_{i})-\mathrm{sinh}(\theta_{i})t))}\, ,\nonumber\\
u^{*}_{i} &=& \frac{N_{i}^{*}\mathrm{cos}(a_{i}) e^{-2i\sqrt{\beta}\mathrm{sin}(a_{i})(\mathrm{cosh}(\theta_{i})t-\mathrm{sinh}(\theta_{i})(x-c_{i}))}}{\mathrm{cosh}(2\sqrt{\beta}\mathrm{cos}(a_{i})(\mathrm{cosh}(\theta_{i})(x-c_{i})-\mathrm{sinh}(\theta_{i})t))}\, ,
\end{eqnarray}
with  $N_{1}=1, \ N_{2}=e^{i\psi}$. As before, the soliton parameters are related to the BT parameters by $a_{i}=A_{i}, \ e^{\theta_{i}} = \delta_{i}$ and the $\alpha_{i}$ from the BT have the form
\begin{eqnarray}\label{eq:alphais}
\alpha_{i} &=&\mathrm{arcsin}\left[\frac{-\mathrm{sin}A_{i}}{\sqrt{1-u_{i}u_{i}^{*}}} \right] \, .
\end{eqnarray}
 The two-soliton solution describes soliton-soliton scattering which will be analysed in the next section.

\subsection{Soliton-soliton scattering}

\begin{figure}[!htp]
\begin{center}
\includegraphics[width=0.75\textwidth]{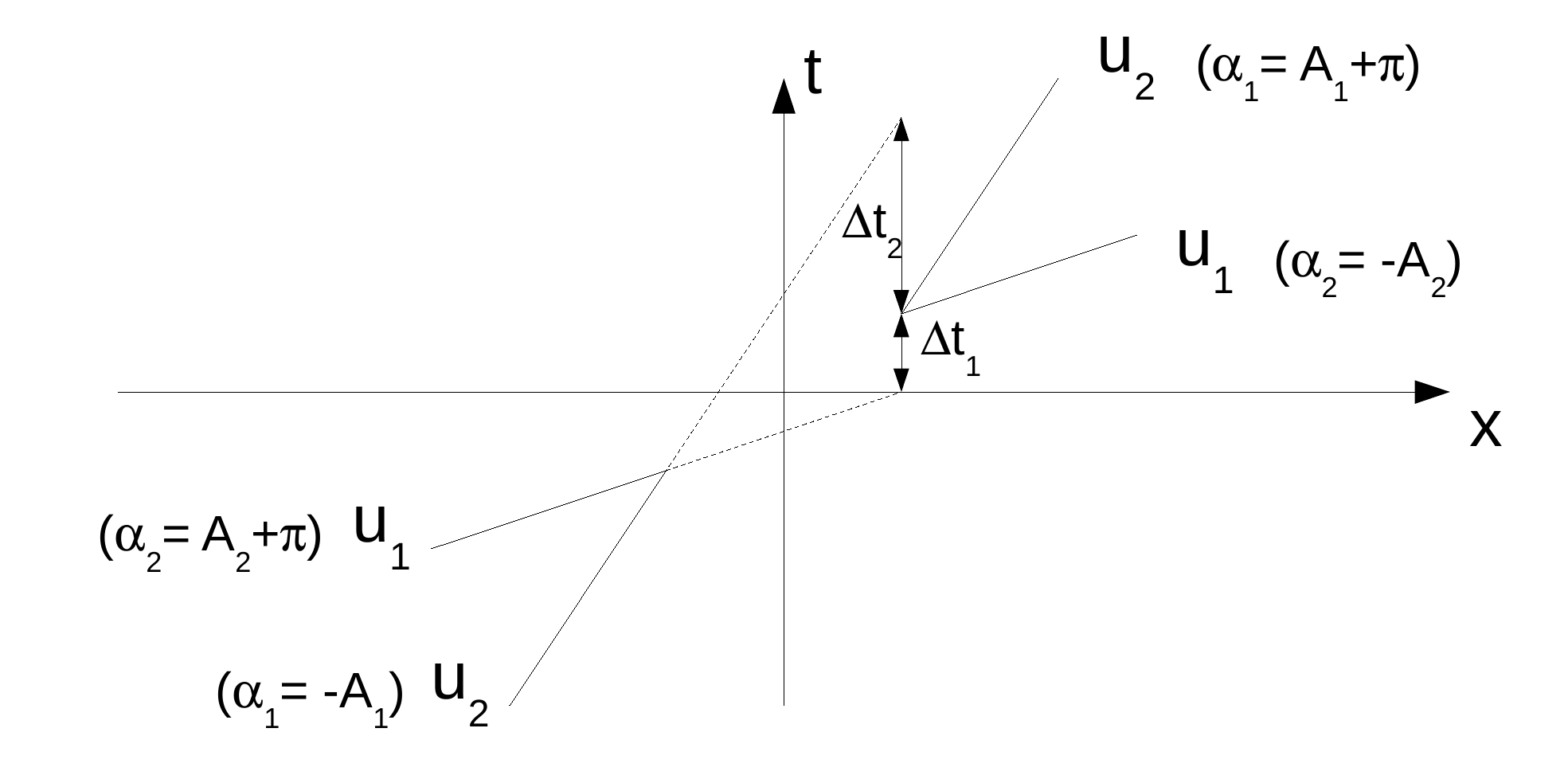}
\end{center}
\caption{Soliton-soliton scattering in general reference frame.}
\label{fig:solsolscat}
\end{figure}

Soliton-soliton scattering is a purely elastic process but the solitons are affected in a subtle way. They experience a time-delay or time-advance and a phase shift. The time-delay and phase shift experienced by two solitons in the scattering process are calculated by analysing the two-soliton solution in particular limits. Namely the form of the scattering solitons needs to be known at the temporal infinities, away from the scattering region. For example to examine soliton $u_{1}$ in the far past one sets $x=\mathrm{tanh}(\theta_{1})t$ and sends $t \rightarrow -\infty$ which sets $u_{2}=0$ and $\alpha_{2}$ to one of its extremal values. The constituent one-soliton solutions $u_{1}$ and $u_{2}$ are generated as previously described, with $\alpha_{i}= A_{i} \pm \pi$ to the left of soliton $u_{i}$ and $\alpha_{i} = -A_{i}$ to the right. Therefore assuming that $\theta_{1} > \theta_{2}$, (as shown in figure \ref{fig:solsolscat}) in the far past $u_{1}$ is to the left of $u_{2}$ so $\alpha_{2} = A_{2}\pm\pi$. The other limits are shown on the diagram.

\begin{figure}[!htp]
\begin{center}
\includegraphics[width=0.8\hsize]{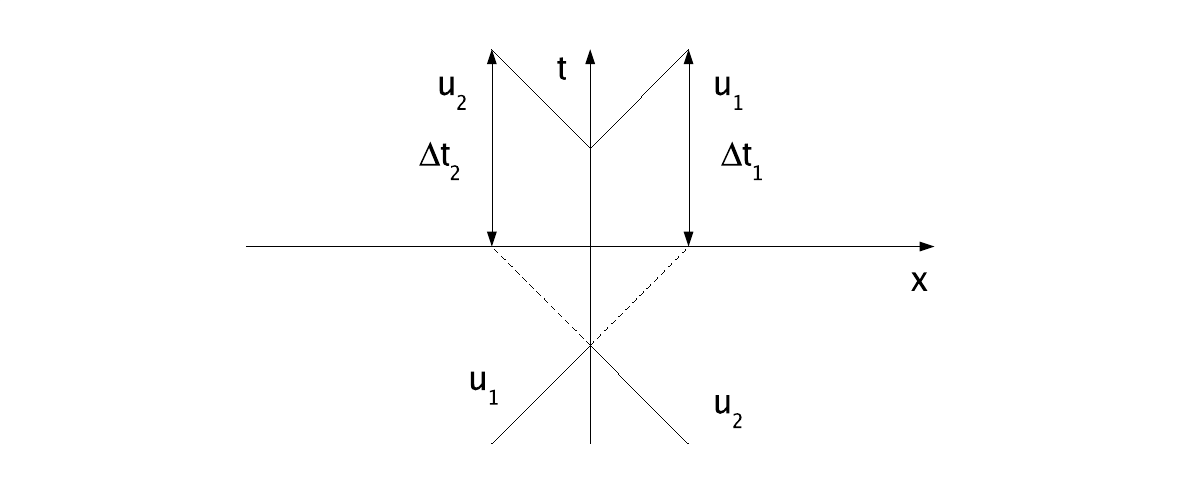}
\end{center}
\caption{Soliton-Soliton scattering in COM frame.}
\label{fig:solsolscatCOM}
\end{figure}

Calculating these limits and comparing with a one-soliton solution with the time shifted by $t \rightarrow t - \Delta t_{i}^{\pm}$ and an extra phase shift $e^{i\phi_{i}^{\pm}}$ gives the time-delays 

\begin{eqnarray}\label{eq:solsolTD}
\Delta t_{1}&=&  \frac{1}{\sqrt{\beta}\mathrm{cos}(a_{1})\mathrm{sinh}(\theta_{1})} \mathrm{ln}\left|\frac{ \mathrm{sinh}\left(\frac{\theta_{1}-\theta_{2}}{2} + i \frac{a_{1}-a_{2}}{2}\right)}{\mathrm{cosh}\left(\frac{\theta_{1}-\theta_{2}}{2} + i \frac{a_{1}+a_{2}}{2}\right)}\right| \, ,\nonumber\\
\Delta t_{2} &=&  \frac{1}{\sqrt{\beta}\mathrm{cos}(a_{2})\mathrm{sinh}(\theta_{2})} \mathrm{ln}\left|\frac{\mathrm{cosh}\left(\frac{\theta_{1}-\theta_{2}}{2} + i \frac{a_{1}+a_{2}}{2}\right)}{ \mathrm{sinh}\left(\frac{\theta_{1}-\theta_{2}}{2} + i \frac{a_{1}-a_{2}}{2}\right)}\right| \, ,
\end{eqnarray}
 and the phase shifts
\begin{eqnarray}\label{eq:solsolphase}
e^{i\phi_{1}} &=& -e^{2\sqrt{\beta} \mathrm{sinh}(\theta_{1}+ia_{1})\Delta t_{1}} \left(\frac{\delta_{2} + \delta_{1} e^{ia_{1}}e^{ia_{2}}}{ \delta_{1} e^{ia_{1}} - \delta_{2} e^{ia_{2}}}\right)^{2} \, ,\nonumber\\
e^{i\phi_{2}}&=& -e^{2\sqrt{\beta} \mathrm{sinh}(\theta_{2}+ia_{2})\Delta t_{2}} \left(\frac{ \delta_{1} e^{ia_{1}} - \delta_{2} e^{ia_{2}}}{\delta_{1} + \delta_{2} e^{ia_{1}}e^{ia_{2}}}\right)^{2}\, .
\end{eqnarray}
There is a special frame of reference, the centre of momentum (COM) frame for which
\begin{equation}
  \sqrt{\beta}\mathrm{cos}(a_{1})\mathrm{sinh}(\theta_{1}) + \sqrt{\beta}\mathrm{cos}(a_{2})\mathrm{sinh}(\theta_{2})=0, 
\end{equation} 
where both solitons are delayed by the same amount,  $\Delta t_{1} = \Delta t_{2} $. This situation is displayed in figure \ref{fig:solsolscatCOM}, in fact this diagram displays the case when $\theta_{1}=-\theta_{2}$ and $a_{1} =a_{2}$. We will now go on and analyse particle-soliton scattering.

\subsection{Particle-soliton scattering}

The particle is taken to be a small fluctuation around the $u=0$ vacuum solution and can also be thought of a one-soliton solution in the small charge limit, i.e $a=\frac{\pi}{2}-\epsilon$. The anti-particle is similarly the one-soliton solution in a different small charge limit, one where the charge is small and negative namely $a = -\frac{\pi}{2} + \epsilon$. In these limits the soliton solution becomes
\begin{eqnarray} \label{eq:particle}
u_{particle} &=& e^{2i\sqrt{\beta}(\mathrm{cosh}(\theta_{p})t - \mathrm{sinh}(\theta_{p})x)}\ , \nonumber\\
u_{antiparticle} &=& e^{-2i\sqrt{\beta}(\mathrm{cosh}(\theta_{p})t - \mathrm{sinh}(\theta_{p})x)}\, .
\end{eqnarray}
The non-topological nature of the soliton sector leads to the unusual property that the small charge soliton and the particle are one and the same thing.

The particle-soliton scattering amplitude is calculated by substituting a small fluctuation $\epsilon e(x,t)$ around the one-soliton solution   
\begin{equation}\label{eq:solpert} u = u_{1-sol}(a,\theta) + \epsilon e(x,t)\, , \end{equation}
into the CSG equation. Solving the differential equation for $e(x,t)$ allows the particle-soliton transmission matrix to be deduced by examining the limits of the solution. When the rapidity of the particle is greater than the rapidity of the soliton the transmission factor is
\begin{eqnarray}\label{eq:transpartsol}
T_{particle/sol} &=& \frac{\mathrm{sinh}\left(\frac{(\theta_{p}-\theta)}{2} + \frac{i}{2}(a-\frac{\pi}{2})\right)^{2}}{\mathrm{sinh}\left(\frac{(\theta_{p}-\theta)}{2} - \frac{i}{2}(a-\frac{\pi}{2})\right)^{2}} \, .
\end{eqnarray}
This result can also be obtained by examining the limit $a_{1}=\frac{\pi}{2}$ in the soliton-soliton scattering phase shift (\ref{eq:solsolphase}). The soliton itself is unaffected during the particle-soliton scattering process.

\section{Complex sine-Gordon theory with defect}
\begin{figure}[htp!]
\begin{center}
\includegraphics[width=0.75\textwidth]{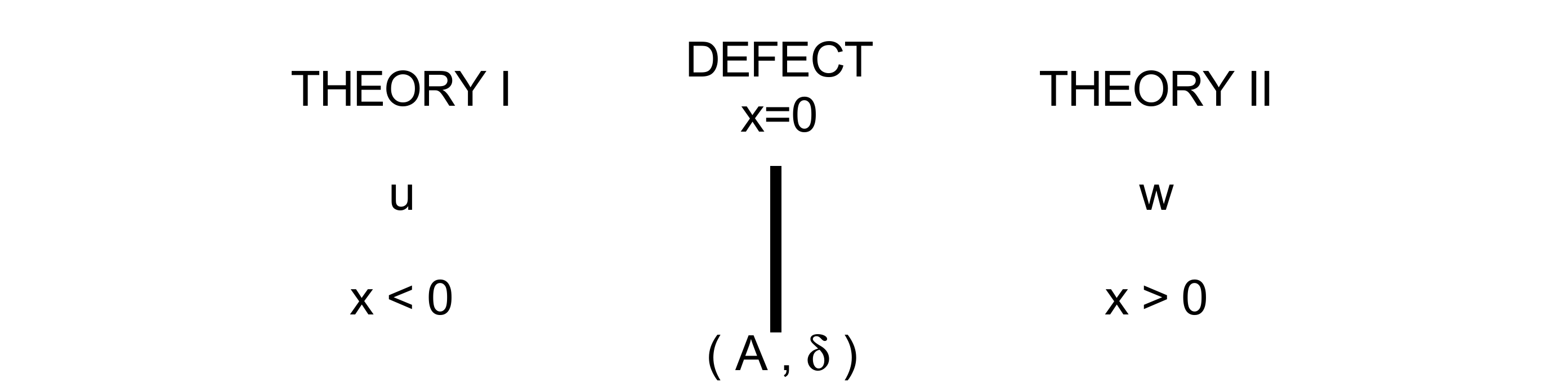}
\end{center}
\caption{Set-up for CSG defect theory.}
\label{Fig:defect}
\end{figure}

In this section we will show that it is possible to sew two CSG theories together with a defect that preserves the classical integrability of the model. Other integrable field theories have been analysed in similar ways \cite{Bowcock:2003dr,Bowcock:2004my,Corrigan:2007gt}. The defect set-up, as shown in figure \ref{Fig:defect}, 
consists of two bulk regions where complex fields $u$ and $w$ satisfy the CSG equation and a defect at $x=0$ where additional constraints (which we shall refer to as defect conditions) couple the fields $u$ and $w$.

\subsection{Constructing the defect theory}

To mathematically construct the defect theory we will assume that the Lagrangian has the following form 

\begin{eqnarray}\label{eq:defectlagrangian}
\mathcal{L} &=& \mathcal{L}^{CSG}_{x<0} +\mathcal{L}^{CSG}_{x>0}   + \left. \mathcal{L}_{D} \right|_{x=0} \nonumber\\
&&\nonumber\\
&=& \mathcal{L}^{CSG}_{x<0} +\mathcal{L}^{CSG}_{x>0}   + \left. \left[ A_{1}u_{t}+ A_{2}u^{*}_{t}+ A_{3}w_{t}+ A_{4}w^{*}_{t} - D(u,w,u^{*},w^{*}) \right] \right|_{x=0} \, . 
\end{eqnarray}
It contains two copies of the standard CSG Lagrangian restricted to the two half-lines and two kinds of defect terms at $x=0$. Any defect terms can be used but to retain the mathematical properties of the bulk theory the terms added here will be chosen to preserve the integrability of the theory. The defect terms that have been used are terms linear in the temporal derivative of the fields, where the coefficients $A_{i} = A_{i}(u,u^{*},w,w^{*})$ are functions of the fields, and a potential term. For simplicity we have excluded the possibility of adding higher time derivatives to the defect Lagrangian, although such terms are 
expected to arise when two or more defects are fused.

By varying the action with respect to the fields in the usual way, the equations of motion and defect conditions are generated. The $u$ and $w$ fields satisfy the CSG equation in their respective regions and satisfy the following defect conditions and their complex conjugate equations at $x=0$:

\begin{eqnarray}\label{eq:DC1}
\frac{\partial_{x} u^{*}}{1-uu^{*}} &=& \frac{\partial \mathcal{L}_{D}}{\partial u} - \frac{\partial}{\partial t} \left( \frac{\partial \mathcal{L}_{D}}{\partial u_{t}}\right) \, , \nonumber \\
-\frac{\partial_{x} w^{*}}{1-ww^{*}} &=& \frac{\partial \mathcal{L}_{D}}{\partial w} - \frac{\partial}{\partial t} \left( \frac{\partial \mathcal{L}_{D}}{\partial w_{t}}\right) \, .
\end{eqnarray}
Explicitly these can be written in terms of $ F_{ij} = \frac{\partial A_{i}}{\partial \phi_j} -\frac{\partial A_{j}}{\partial \phi_i} $ and $ {\vec\phi}=(u,u^{*},w,w^{*})$  as
\begin{eqnarray}\label{eq:DC}
\frac{\partial_{x} u^{*}}{1-uu^{*}} &=& (F_{21}u^{*}_{t} +F_{31}w_{t}+F_{41}w^{*}_{t}) - \frac{\partial D}{\partial u}  \, , \nonumber \\
-\frac{\partial_{x} w^{*}}{1-ww^{*}} &=& (F_{13}u_{t} +F_{23}u^{*}_{t}+F_{43}w^{*}_{t})   - \frac{\partial D}{\partial w} \, .
\end{eqnarray}
The defect conditions are first order equations in the left and right fields. 
Recalling that the B\"{a}cklund transformation (\ref{eq:DefectCond}) also has this form and 
following the example of previous studies into integrable defect theories, it is expected that 
the defect conditions that preserve integrability are given by the BT. 
Writing the BT in the same form as the defect conditions,
\begin{equation}\label{eq:Backlund}
\frac{\partial_{x} u^{*}}{1-uu^{*}} = \hdots , \ \ -\frac{\partial_{x} w^{*}}{1-ww^{*}} = \hdots \, ,
\end{equation}
allows the two to be compared  and the unknown functions in the Lagrangian, $D(u,u^{*},w,w^{*})$  and $A_{i}$'s to be solved for, yielding
\begin{eqnarray}\label{eq:D}
D & =& \delta\sqrt{\beta} \left(2\ \mathrm{cos}(\alpha)\ \sqrt{1-uu^{*}}\sqrt{1-ww^{*}}-uw^{*}-u^{*}w\right) \nonumber\\
&&+ \frac{\sqrt{\beta}}{\delta}\left(2\ \mathrm{cos}(\alpha)\ \sqrt{1-uu^{*}}\sqrt{1-ww^{*}}+uw^{*}+u^{*}w\right) \, , 
\end{eqnarray}

\begin{eqnarray}\label{eq:Ais}
A_{1} = A_{2}^{*}&=& \frac{1}{u}\mathrm{ln}\left(2\sqrt{1-uu^{*}}\sqrt{1-ww^{*}}e^{-i\alpha}-2(1-uu^{*})\frac{w}{u}\right)-\frac{1}{2} \frac{\mathrm{ln}(w)}{u}\, ,\nonumber\\
A_{3} = A_{4}^{*}&=& -\frac{1}{w}\mathrm{ln}\left(2\sqrt{1-uu^{*}}\sqrt{1-ww^{*}}e^{i\alpha}-2(1-ww^{*})\frac{u}{w}\right)+\frac{1}{2} \frac{\mathrm{ln}(u)}{w} \, .\nonumber\\
&&
\end{eqnarray}
Note that this solution is not unique; we are free to make gauge transformations of the form
$A_{i}\rightarrow A_{i} + \partial\chi/\partial\phi_{i}$. Such transformations change the Lagrangian
by a total derivative and do not affect the equations of motion.

We now turn to the question of whether the defect Lagrangian we have constructed, using defect conditions derived from the B\"{a}cklund transformation, preserves the integrability of the model. This can be checked by explicitly constructing formulae for conserved energy, momentum and charge. The existence of these three conserved charges is not enough in itself to ensure integrability, but  previous work suggests that if a conserved energy and a conserved momentum exist, then conservation of higher spin charges follows. The energy in the presence of a defect can be read directly from the Lagrangian,
\begin{equation}
 E_{def} = \int^{0}_{-\infty} \mathrm{d}x \ \mathcal{E}^{CSG}(u,u^{*}) + \int_{0}^{\infty} \mathrm{d}x \ \mathcal{E}^{CSG}(w,w^{*}) + \left. D(u,u^{*},w,w^{*}) \right|_{x=0} 
\end{equation} 
and its conservation $\frac{\partial E_{def}}{\partial t}=0$ can be checked. Unlike the defect energy, the defect momentum and charge cannot be written down from the Lagrangian. To find these we ask what term can be added to the usual bulk quantities to create a quantity conserved in the defect theory

\begin{eqnarray}\label{eq:MomCharge}
 P_{def} &=& \int^{0}_{-\infty} \mathrm{d}x \ \mathcal{P}^{CSG}(u,u^{*}) + \int_{0}^{\infty} \mathrm{d}x \ \mathcal{P}^{CSG}(w,w^{*}) + \left. \mathcal{P}_{D}(u,u^{*},w,w^{*}) \right|_{x=0} \, , \nonumber \\
 Q_{def} &=& \int^{0}_{-\infty} \mathrm{d}x \ \mathcal{Q}^{CSG}(u,u^{*}) + \int_{0}^{\infty} \mathrm{d}x \ \mathcal{Q}^{CSG}(w,w^{*}) + \left. \mathcal{Q}_{D}(u,u^{*},w,w^{*}) \right|_{x=0} \, .
\end{eqnarray} 
 At first look it does not seem that any $\mathcal{P}_{D}$ would allow the defect momentum to be conserved because the defect breaks the translational invariance of the theory. In fact it is possible to find a conserved defect momentum because the defect conditions hold true at all points and not just at $x=0$. By demanding that $\frac{\partial P_{def}}{\partial t} =0$ and  $\frac{\partial Q_{def}}{\partial t} =0$, the following defect terms are derived 

\begin{eqnarray}
\mathcal{P}_{D} &=& -  \delta \sqrt{\beta} \left(-2\ \mathrm{cos}(\alpha)\ \sqrt{1-uu^{*}}\sqrt{1-ww^{*}} + u^{*}w+uw^{*}\right)\nonumber\\
&& - \frac{\sqrt{\beta}}{\delta}\left(2\ \mathrm{cos}(\alpha)\ \sqrt{1-uu^{*}}\sqrt{1-ww^{*}} + u^{*}w+w^{*}u\right) \, ,\nonumber \\
\mathcal{Q}_{D} &=&2\alpha \, .
\end{eqnarray}
It can be checked that the defect energy and momentum constructed here satisfy the usual relation
\begin{equation}
\frac{\partial E_{def}}{\partial t} + \frac{\partial P_{def}}{\partial x} = 0.
\end{equation}

\subsection{Soliton solutions in the defect theory}

In this section we will look for soliton solutions that solve the defect conditions. We shall consider a number of scenarios using different numbers of soliton solutions on each side of the defect. Solutions with one soliton on each side of the defect will be used to study soliton-defect scattering; solutions with a single soliton on one side of the defect and the vacuum solution on the other will be used to examine the possibility of the defect absorbing or emitting a soliton.

Firstly, note that the bulk vacuum on each side of the defect $(u=0,w=0)$ trivially solves the defect conditions. In this case the formulae for the energy, momentum and charge receive no contribution from the bulk but a non-zero contribution at $x=0$ which we attribute to  the defect. Explicitly these are given by
\begin{eqnarray}\label{eq:defectvacuum}
E_{vac} &=& 2\sqrt{\beta}\ (\delta+\frac{1}{\delta})\ \mathrm{cos}(\alpha_0) \, ,\nonumber\\
P_{vac} &=& 2\sqrt{\beta}\ (\delta-\frac{1}{\delta})\ \mathrm{cos}(\alpha_0)\, ,\nonumber\\
Q_{vac} &=& 2 \alpha_0 \, ,
\end{eqnarray}
where $\alpha_0 =\alpha(u=0,w=0)$.  These expressions depend on the two defect parameters $\delta$ and $A$, through $\alpha_0$. The non-trivial way in which $\alpha_0$ is related to $A$  

\begin{equation}\label{eq:alphavacuum} \mathrm{sin}(\alpha_0)= -\mathrm{sin}(A) \, ,\end{equation}
leads to the interesting property that the energy, momentum and charge can take different values despite the bulk fields being the same $u=0,\ w=0$. Explicitly (\ref{eq:alphavacuum}) has the solutions \begin{equation}\alpha_0 = -A,\ A + \pi,\end{equation} where the two solutions give different values to $\mathrm{cos}(\alpha_0)$ and therefore to the energy and momentum. In particular there are defects with positive energy and defects with negative energy (with either positive or negative momentum), which we will refer to as excited and unexcited defects respectively.

There are four excited defects (i.e. with positive energy) with the same energy and momentum but with different charge, where we take $ \mathrm{cos}(\alpha_0) = \mathrm{cos}(A) $ (taking the other solution just gives another cover of the solutions). These excited defects are displayed in table \ref{tab:exdef}, for the example when $A'=\frac{\pi}{3},\ \delta' = 2$ the defects have energy, $ E_{def} = \frac{5\sqrt{\beta}}{2} $ and  momentum, $  P_{def} = \frac{3\sqrt{\beta}}{2} $ with charges, $Q_{def} = \left\{\pm \frac{2\pi}{3}, \ \pm \frac{4\pi}{3} \right\}$. 
\begin{table}[h]
\begin{center}
\fbox{
\begin{tabular}{rcl|l|ll|ll}
$0 <$&$ A$&$ < \frac{\pi}{2}$ &$ A = A'$ &$ \delta =\delta'$&$> 0$&$ Q =-2A' $&$\in \left\{ 0,-\pi \right\}$ \\
$- \frac{\pi}{2} <$&$ A$&$ < 0$ &$ A = -A'$&$ \delta =\delta'$&$> 0 $&$ Q =2A' $&$\in \left\{ \pi,0 \right\}$ \\
$\frac{\pi}{2} <$&$ A$&$ < \pi $ &$ A =\pi -A'$&$ \delta =-\delta'$&$< 0$&$ Q =2A' - 2\pi $&$\in \left\{ -\pi,-2\pi \right\} $ \\
$-\pi <$&$ A$&$ < -\frac{\pi}{2}$ &$ A =-\pi +A'$&$ \delta=-\delta'$&$ < 0$ &$ Q =2\pi - 2A' $&$\in \left\{ 2\pi,\pi \right\} $

\end{tabular}
}
\end{center}
\caption{Table showing four excited defects with the same energy and momentum.}
\label{tab:exdef}
\end{table}

Similarly there are four unexcited defects, shown in table \ref{tab:undef}, with the same negative energy and momentum (positive or negative) but with different charge, again $ \mathrm{cos}(\alpha_0) = \mathrm{cos}(A) $ is used in the calculations. Taking $A' = \frac{\pi}{3}\ ,\  \delta' = -2$ gives four defects with energy, $ E_{def} = -\frac{5\sqrt{\beta}}{2} $ and  momentum, $  P_{def} = \frac{3\sqrt{\beta}}{2} $ with charges $Q_{def} = \left\{\pm \frac{2\pi}{3}, \ \pm \frac{4\pi}{3} \right\}$. 

\begin{table}[h]
\begin{center}
\fbox{
\begin{tabular}{rcl|l|ll|ll}
$0 < $&$A$&$ < \frac{\pi}{2}$ &$A=A'$&$ \delta =\delta'$&$< 0$ &$ Q =-2A'$&$\in \left\{ 0,-\pi \right\}$ \\
$- \frac{\pi}{2} <$& $A$&$ < 0$ &$A=-A'$&$ \delta =\delta'$&$< 0 $&$ Q =2A'$&$\in \left\{ \pi,0 \right\}$  \\
$\frac{\pi}{2} <$& $A$&$ < \pi $&$A=\pi-A'$&$ \delta =-\delta'$&$> 0$ &$ Q =2A'-2\pi$&$\in \left\{ -\pi,-2\pi \right\} $ \\
$-\pi <$& $A$&$ < -\frac{\pi}{2}$ &$A = -\pi + A'$&$ \delta =-\delta'$&$> 0$ &$ Q =2\pi-2A'$&$\in \left\{ 2\pi,\pi \right\} $
\end{tabular}
}
\end{center}
\caption{Table showing four unexcited defects with the same energy and momentum.}
\label{tab:undef}
\end{table}

\subsubsection{Soliton emission}

Na\"ively one might think that defects are unable to absorb or emit solitons in the CSG model, since at 
early/late times the bulk configuration would look like the vacuum near the defect, and one would still have to account for the conserved charges carried by the absorbed or emitted soliton. However,
the findings in the previous section show that though the bulk fields are in their vacuum configuration, the defect can have different values of energy, momentum and charge depending on the two defect parameters $\delta$ and $A$. Specifically it points towards an unexcited defect absorbing a soliton, thus exciting it and an excited defect decaying into an unexcited defect by the emission of a soliton. This picture is further suggested by the fact that the energy of a soliton (\ref{eq:energysol}) is exactly twice the energy of a defect (\ref{eq:defectvacuum}), when the relevant parameters are matched, and defects of the positive and negative energies of the same magnitude are linked by the solutions to equation (\ref{eq:alphavacuum}). Since the defect conditions and the B\"{a}cklund transformation are one and the same thing the properties of the BT also provide evidence for the absorption and emission of solitons by the defect. We have already seen that the BT provides a means of generating a one-soliton solution from the vacuum, and the emission of a soliton from the defect is exactly this situation with the process happening dynamically.

To examine the emission of a soliton we need to analyse the field configurations to replicate a right moving soliton being emitted from the right hand side of the defect. The set-up required is the null field on the left hand side of the defect and a right moving one-soliton solution on the right. In this situation $\alpha$ becomes
\begin{equation} \alpha' = \mathrm{arcsin}\left( \frac{-\mathrm{sin}(A)}{\sqrt{1-ww^{*}}}\right)\, , \end{equation}
where $w=w_{1-sol}$ is the specified one-soliton solution. In this scenario the defect conditions (\ref{eq:DefectCond}) are satisfied, if $\mathrm{cos}(\alpha') =\mathrm{cos}(A)$ at $t \rightarrow -\infty$, by \begin{eqnarray} a= A &,&e^{\theta} = \delta \nonumber\, ,\\a = A \pm \pi &,&e^{\theta} = -\delta \, .\end{eqnarray}
Since the emitted soliton is a right moving soliton, $\theta > 0$, the above conditions on the defect parameters show that for a defect to emit a right moving soliton then $|\delta| > 1$ is required. This is explained by a right moving soliton having positive energy and momentum meaning that the initial excited defect has to have positive energy and momentum for total energy and total momentum to be conserved.

\begin{table}[!h]
\begin{center}
\begin{tabular}{|ccccc|}
\hline
Excited Defect I && $\rightarrow$& & Unexcited Defect I \\
&&&&\\
$( 0 < A < \frac{\pi}{2} \ , \ \delta > 1)$ &&Emitted Soliton&&$ ( \frac{\pi}{2} < A' < \pi \ , \  \delta' > 1)$ \\
$\alpha = -A $&&$(a=A,e^{\theta}=\delta)$&&$ \alpha = A - \pi $\\
$E = 2\sqrt{\beta}\mathrm{cos}(A)(\delta+\frac{1}{\delta})$ &&$E_{sol}=4\sqrt{\beta}\mathrm{cos}(A)(\delta+\frac{1}{\delta})$&&$ E' = -E$\\
$P = 2\sqrt{\beta}\mathrm{cos}(A)(\delta-\frac{1}{\delta})$ &&$P_{sol}=4\sqrt{\beta}\mathrm{cos}(A)(\delta-\frac{1}{\delta})$&&$ P' = -P$\\ 
$Q = -2A $&&$Q_{sol}=2\pi - 4A$&&$ Q' = 2A -2\pi$ \\
\hline
\end{tabular} 
\end{center}
\caption{Table showing the decay process of excited defect I.}\label{fig:Emission1}
\end{table}
Examining the decays of two of the four excited defects, the first shown in table \ref{fig:Emission1}, where excited defect I with charge in the region, $-\pi < Q_{def} < 0$ decays by emitting a positively charged soliton. Similarly table \ref{fig:Emission2} shows excited defect II with charge, $\pi < Q_{def} < 2\pi$ decaying by emitting a positively charged soliton. 

\begin{figure}[!htp]
\begin{center}
\subfigure[]{\label{fig:alphaA}\includegraphics[width=0.25\textwidth]{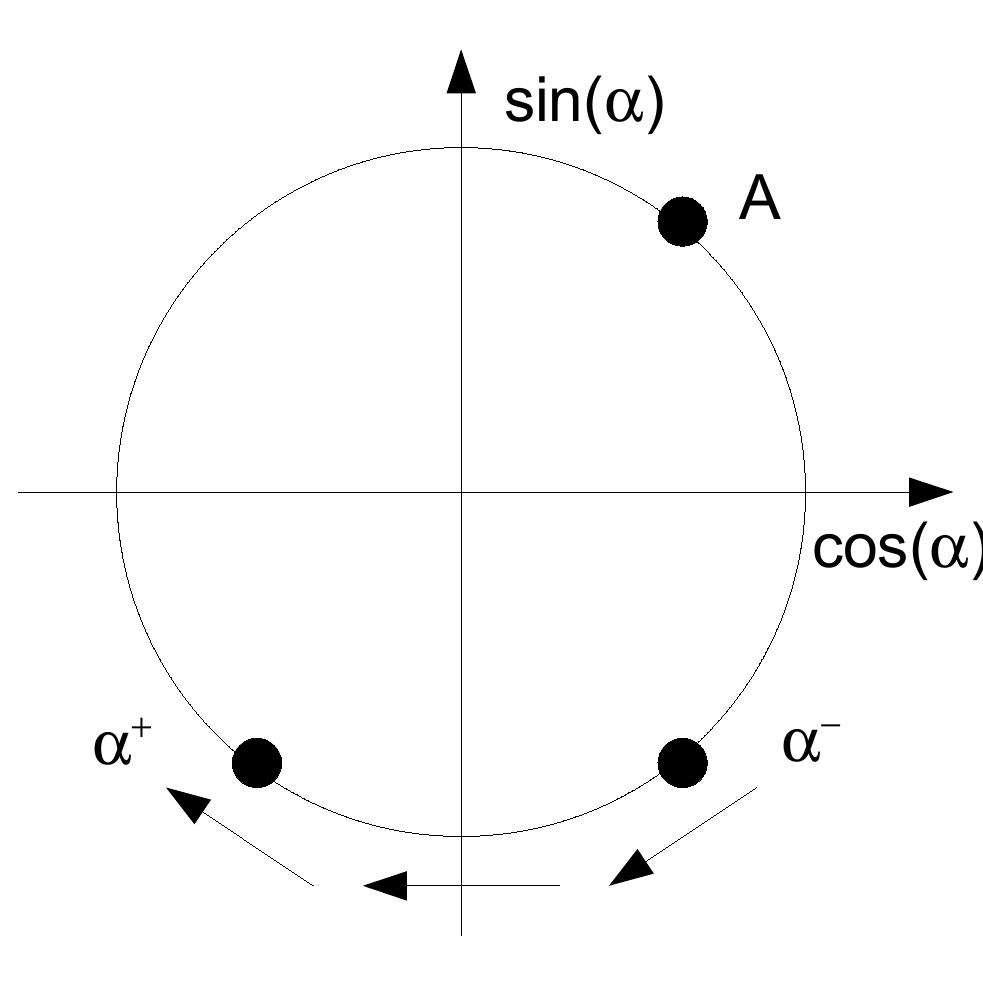}}
\hspace{0.8in} 
\subfigure[]{\label{fig:alphaB}\includegraphics[width=0.25\textwidth]{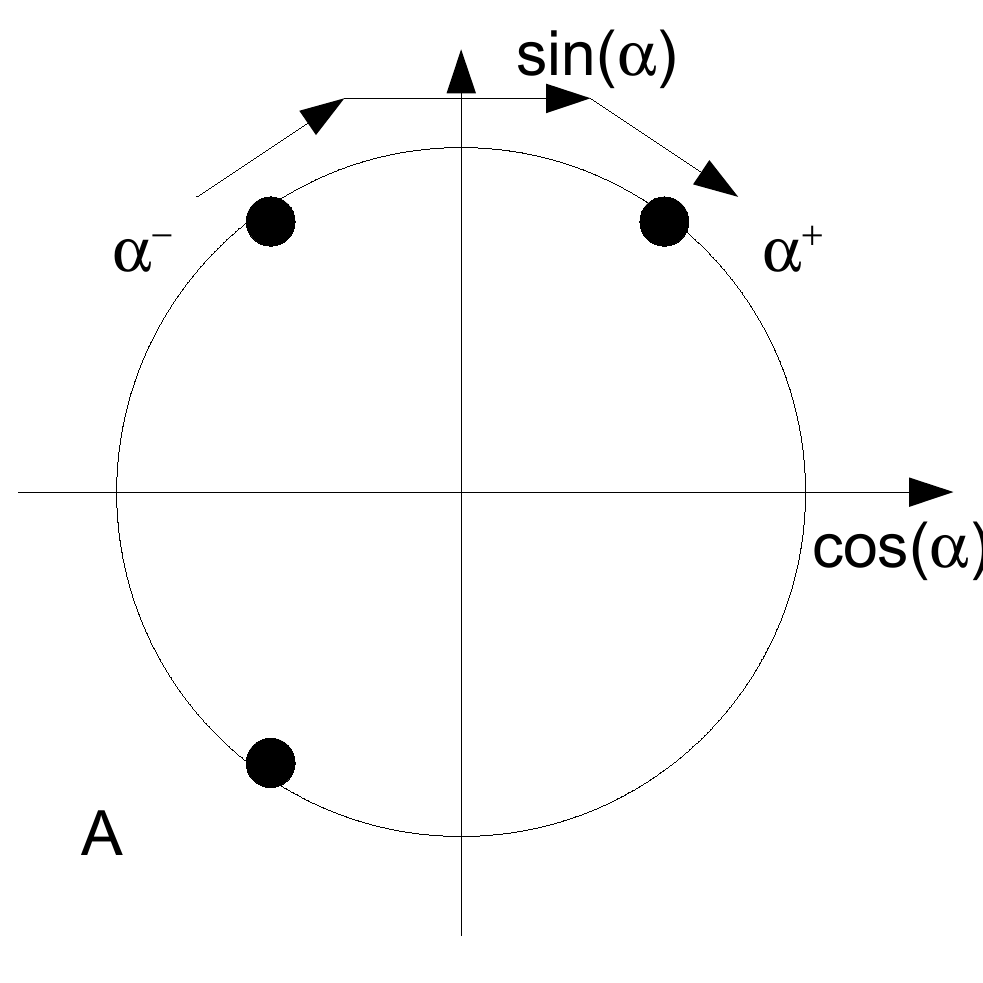}}
\end{center}
\caption{Graphic showing the evolution of $e^{i\alpha'}$ during two emission processes.}
\label{fig:alpha}
\end{figure}

\begin{table}[!h]
\begin{center}
\begin{tabular}{|ccccc|}
\hline
Excited Defect II &&$ \rightarrow$ && Unexcited Defect II \\
&&&&\\
$( -\pi < A < -\frac{\pi}{2} \ ,\  \delta < -1   )$ &&Emitted Soliton&&$ ( 0 > A' > -\frac{\pi}{2} \ , \ \delta' < -1   )$ \\
$\alpha = -A $&&$(A-\pi,-\delta)$&&$ \alpha = A + \pi $\\
$E = 2\sqrt{\beta}\mathrm{cos}(A)(\delta+\frac{1}{\delta})$ &&$E_{sol}=4\sqrt{\beta}\mathrm{cos}(A)(\delta+\frac{1}{\delta})$&&$ E' = -E$\\
$P = 2\sqrt{\beta}\mathrm{cos}(A)(\delta-\frac{1}{\delta})$ &&$P_{sol}=4\sqrt{\beta}\mathrm{cos}(A)(\delta-\frac{1}{\delta})$&&$ P' = -P$\\ 
$Q = -2A$ &&$Q_{sol}=-2\pi-4A$&&$ Q' = 2A +2\pi$ \\
\hline
\end{tabular} 
\end{center}
\caption{Table showing the decay process of excited defect II.}\label{fig:Emission2}
\end{table}
The two decays illustrate that each defect with positive energy and momentum emits precisely one right moving soliton. During any emission process the value of $\alpha'$ changes with time from one solution of (\ref{eq:alphavacuum}) to the other. Without loss of generality we take $\alpha' = -A$ in the far past before the soliton emission has occurred. Due to the way that $\alpha$ is defined 
\begin{equation}
\cos(\alpha') =\  \sqrt{1- \sin^2(\alpha')} =\ \frac{\sqrt{1-ww^{*} -\sin^2(A)}}{\sqrt{1-ww^{*}}} \, ,
\end{equation}
is only defined using a square root. During the process 
\begin{equation} \sqrt{1-ww^{*}} = \sqrt{1 - \mathrm{cos}^{2}(a)\mathrm{sech}^{2}(2\sqrt{\beta}\mathrm{cos}(a)(-\mathrm{sinh}(\theta)t))}  \end{equation}
becomes equal to $\mathrm{sin}(A)$, at this moment the argument in the squareroot in the definition of $\cos(\alpha')$ becomes equal to zero and the opposite branch of this squareroot is chosen to keep the function smooth. Using this prescription, by the end of the process in the far future $\alpha'$ is equal to the other solution of (\ref{eq:alphavacuum}), namely $\alpha' = A \pm \pi$. This evolution is shown in figure \ref{fig:alpha} for the two decays illustrated in tables \ref{fig:Emission1} and \ref{fig:Emission2}. 
The figure shows the value of $e^{i\alpha'}$ in the complex plane changing during both emissions from $\alpha' = -A$ to $\alpha' = A \pm \pi$. Figure \ref{fig:alphaA} shows the evolution for $e^{i\alpha'}$ during the decay of excited defect I where $0 < A < \frac{\pi}{2}$ and figure \ref{fig:alphaB} illustrates the progress of $e^{i\alpha'}$ during the decay of excited defect II where  $-\pi < A < -\frac{\pi}{2}$. It is noted that in both cases the value of $\mathrm{sin}(\alpha')$ is the same at the beginning and the end, while the value of $\mathrm{cos}(\alpha')$ is different. 

As mentioned earlier a specific defect emits just one type of soliton but there is the interesting phenomena that there are two defects that emit the same soliton. The decay of two such defects are shown in tables \ref{fig:emission1} and \ref{fig:emission2}. The two excited defects, $E_{A}$ and $E_{B}$, have the same energy and momentum but different charge, $Q = -\frac{2\pi}{3}$ and $Q = \frac{4\pi}{3}$ respectively and they emit the same soliton, specifically a soliton with energy, $E_{sol}=5\sqrt{\beta}$, momentum, $P_{sol}=3\sqrt{\beta}$ and charge, $Q_{sol}=\frac{2\pi}{3}$.  
\begin{table}[!h]
\begin{center}
\begin{tabular}{|ccccc|}
\hline
$E_{A}$ &&$ \rightarrow$& &$ U_{A} $\\
&&&&\\
$(  A = \frac{\pi}{3} \ , \ \delta = 2)$ &&Emitted Soliton&&$ ( A' =\frac{2\pi}{3} \ , \  \delta' = 2)$ \\
$\alpha = -A $&&$(a=A,e^{\theta}=\delta)$&&$ \alpha = A - \pi $\\
$E = \frac{5\sqrt{\beta}}{2}$ &&$E_{sol}=5\sqrt{\beta}$&&$ E' = -E$\\
$P = \frac{3\sqrt{\beta}}{2}$ &&$P_{sol}=3\sqrt{\beta}$&&$ P' = -P$\\ 
$Q = -\frac{2\pi}{3}$ &&$Q_{sol}=\frac{2\pi}{3}$&&$ Q' = -\frac{4\pi}{3}$ \\
\hline
\end{tabular}
\end{center}
\caption{Table showing the decay of excited defect $E_{A}$.}
\label{fig:emission1}
\end{table}
\begin{table}[!h]
\begin{center}
\begin{tabular}{|ccccc|}
\hline
$E_{B}$ && $\rightarrow$ &&$ U_{B}$ \\
&&&&\\
$( A =-\frac{2\pi}{3} \ ,\  \delta = -2   )$ &&Emitted Soliton&&$ (  A' = -\frac{\pi}{3} \ , \ \delta' = -2   )$ \\
$\alpha = -A $&&$(A-\pi,-\delta)$&&$ \alpha = A + \pi $\\
$E =  \frac{5\sqrt{\beta}}{2}$ &&$E_{sol}=5\sqrt{\beta}$&&$ E' = -E$\\
$P = \frac{3\sqrt{\beta}}{2} $ &&$P_{sol}=3\sqrt{\beta}$&&$ P' = -P$\\ 
$Q = \frac{4\pi}{3}$ &&$Q_{sol}=\frac{2\pi}{3}$&& $Q' =\frac{ 2\pi}{3}$ \\
\hline
\end{tabular}
\end{center}
\caption{Table showing the decay of excited defect $E_{B}$.}
\label{fig:emission2}
\end{table}

\begin{figure}[!!h]
\begin{center}
\includegraphics[width=0.55\hsize]{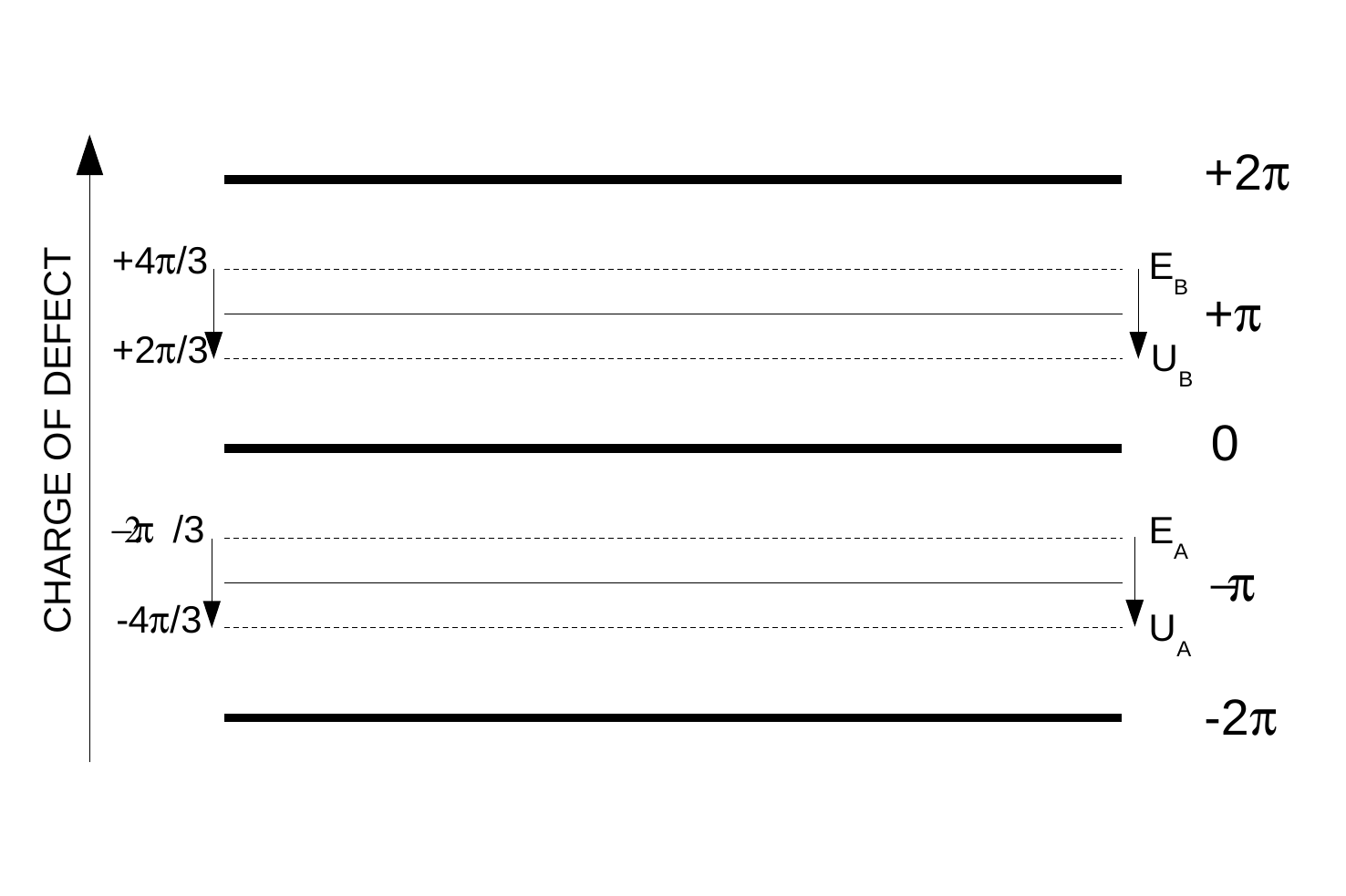}
\end{center}
\caption{Graphic showing charge of defects $E_A$ and $E_B$ that emit a $Q=\frac{2\pi}{3}$ soliton.}
\label{fig:defectcharge}
\end{figure}

The change of charge from the initial excited defect to the final unexcited defect for the two processes are graphically illustrated in figure \ref{fig:defectcharge}.
It shows that the initial and final charge of the defects are symmetrically spaced about $Q = \pm \pi$ and this is the case for any emission. The maximally charged soliton $Q_{sol} = 2\pi$ is emitted by defects with charge $2\pi$ or $0$ while the defects with charge $\pm \pi$ are stable and do not emit any soliton. For any emission the charge of the unexcited defect is never the opposite of the initial charge. Considering the pair of defects that emit the same soliton (for example $E_A$ and $E_B$) then the unexcited defect that the other excited defect of the pair decays into does has the opposite charge of the first excited defect, as well as energy and momentum. In this example $U_B$ has the opposite charge to $E_A$, as do $U_A$ and $E_B$. We will call this the anti-defect, for example $U_{B}$ is the anti-defect of $E_{A}$ and vice-versa.

\subsubsection{Soliton absorption}
The opposite process to a soliton being emitted from the defect is for a soliton to be absorbed by an unexcited defect. The field configurations needed to describe this process are a right moving one-soliton solution on the left hand side of the defect and the null solution on the right. In this situation $\alpha$ becomes \begin{equation} \alpha' = \mathrm{arcsin}\left( \frac{\mathrm{sin}(A)}{\sqrt{1-uu^{*}}}  \right) \, ,\end{equation} where $u=u_{1-sol}$ is a right moving one-soliton. The defect conditions (\ref{eq:DefectCond}) are satisfied, again using $\mathrm{cos}(\alpha')= \mathrm{cos}(A)$ as $t \rightarrow \infty$, when \begin{eqnarray}  a = -A&,&  e^{\theta} = -\delta\, ,\nonumber\\ a = -A \pm \pi &,& e^{\theta} = \delta \, .\end{eqnarray}
These conditions show that it is required that $|\delta| > 1$ for a defect to absorb a right moving soliton. This can be explained by noticing that for an unexcited defect to absorb a right moving soliton is has to have negative energy and momentum which requires $|\delta| > 1$.
\begin{table}[!h]
\begin{center}
\begin{tabular}{|ccccc|}
\hline
Unexcited Defect I && $\rightarrow$& & Excited Defect I \\
&&&&\\
$( \frac{\pi}{2} < A < \pi \ , \ \delta > 1)$ &&Absorbed Soliton&&$ ( 0 < A' < \frac{\pi}{2}  , \  \delta' > 1)$ \\
$\alpha = -A $&&$(a=-A+\pi,e^{\theta}=\delta)$&&$ \alpha = A - \pi $\\
$E = 2\sqrt{\beta}\mathrm{cos}(A)(\delta+\frac{1}{\delta})$ &&$E_{sol}=-4\sqrt{\beta}\mathrm{cos}(A)(\delta+\frac{1}{\delta})$&&$ E' = -E$\\
$P = 2\sqrt{\beta}\mathrm{cos}(A)(\delta-\frac{1}{\delta})$ &&$P_{sol}=-4\sqrt{\beta}\mathrm{cos}(A)(\delta-\frac{1}{\delta})$&&$ P' = -P$\\ 
$Q = -2A$ &&$Q_{sol}=-2\pi + 4A$&&$ Q' = 2A -2\pi $\\
\hline
\end{tabular}
\end{center}
\caption{Table showing  unexcited defect I absorbing a soliton.}
\label{fig:absorption}
\end{table}
\begin{table}[!h]
\begin{center}
\begin{tabular}{|ccccc|}
\hline
Unexcited Defect II &&$ \rightarrow$ && Excited Defect II \\
&&&&\\
$(  -\frac{\pi}{2} < A < 0 \ ,\  \delta < -1   )$ &&Absorbed Soliton&&$ ( -\pi < A' < -\frac{\pi}{2} \ , \ \delta' < -1   )$ \\
$\alpha = -A $&&$(-A,-\delta)$&&$ \alpha = A + \pi $\\
$E = 2\sqrt{\beta}\mathrm{cos}(A)(\delta+\frac{1}{\delta})$ &&$E_{sol}=-4\sqrt{\beta}\mathrm{cos}(A)(\delta+\frac{1}{\delta})$&&$ E' = -E$\\
$P = 2\sqrt{\beta}\mathrm{cos}(A)(\delta-\frac{1}{\delta})$ &&$P_{sol}=-4\sqrt{\beta}\mathrm{cos}(A)(\delta-\frac{1}{\delta})$&&$ P' = -P$\\ 
$Q = -2A$ &&$Q_{sol}=2\pi+4A$&&$ Q' = 2A +2\pi$ \\
\hline
\end{tabular}
\end{center}
\caption{Table showing unexcited defect II absorbing a soliton.}
\label{fig:absorption1}
\end{table}

Table \ref{fig:absorption} shows that unexcited defect I described by $\frac{\pi}{2} < A < \pi$ with $Q \in \left\{-\pi,\ -2\pi \right\}$ absorbs a positively charged soliton becoming excited defect I. Similarly table \ref{fig:absorption1} shows that unexcited defect II described by $-\frac{\pi}{2} < A < 0$ with $Q \in \left\{0, \ \pi \right\}$ absorbs a positively charged soliton becoming excited defect II. In both absorption processes shown in the tables it can be checked that the total energy, momentum and charge are conserved, for example the energy of the unexcited defect plus the energy of the absorbed soliton is equal to the energy of the excited defect.

\begin{figure}[!h]
\begin{center}
\subfigure[]{\label{fig:alpha2A}\includegraphics[width=0.25\textwidth]{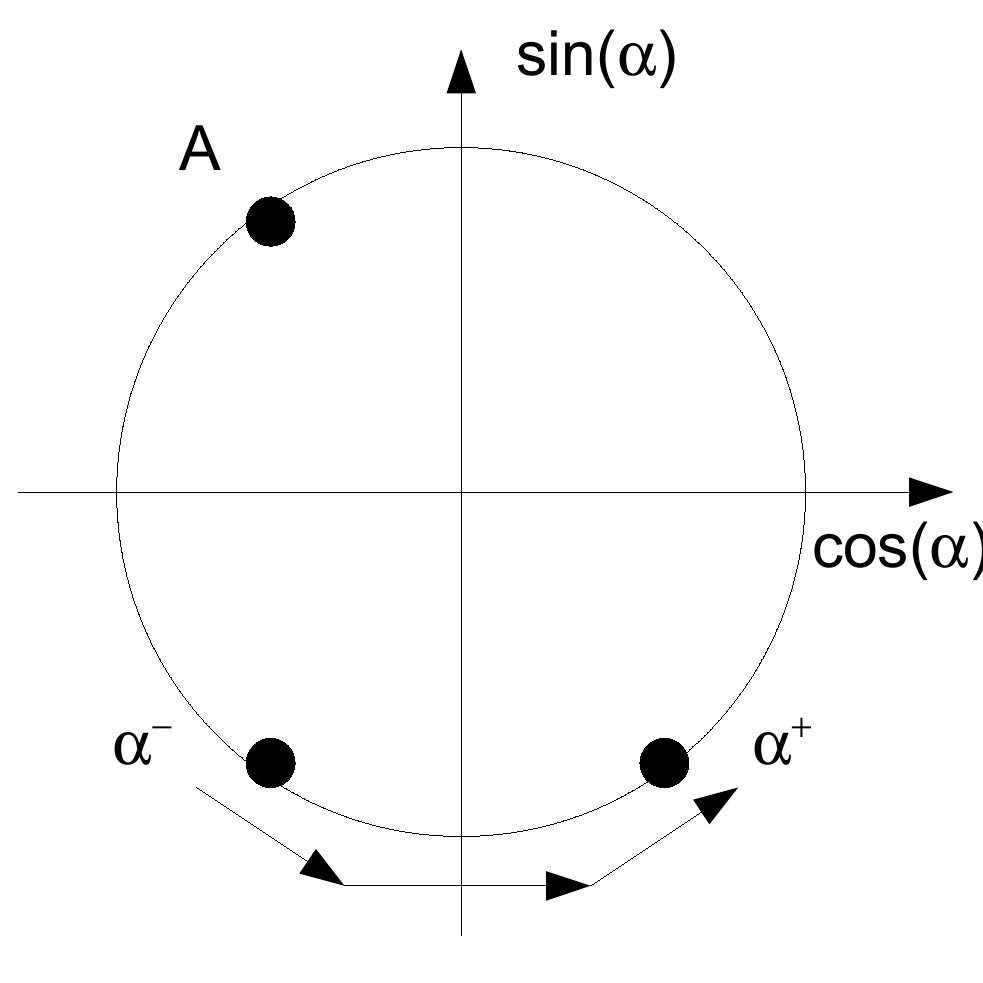}}
\hspace{0.8in}
\subfigure[]{\label{fig:alpha2B}\includegraphics[width=0.25\textwidth]{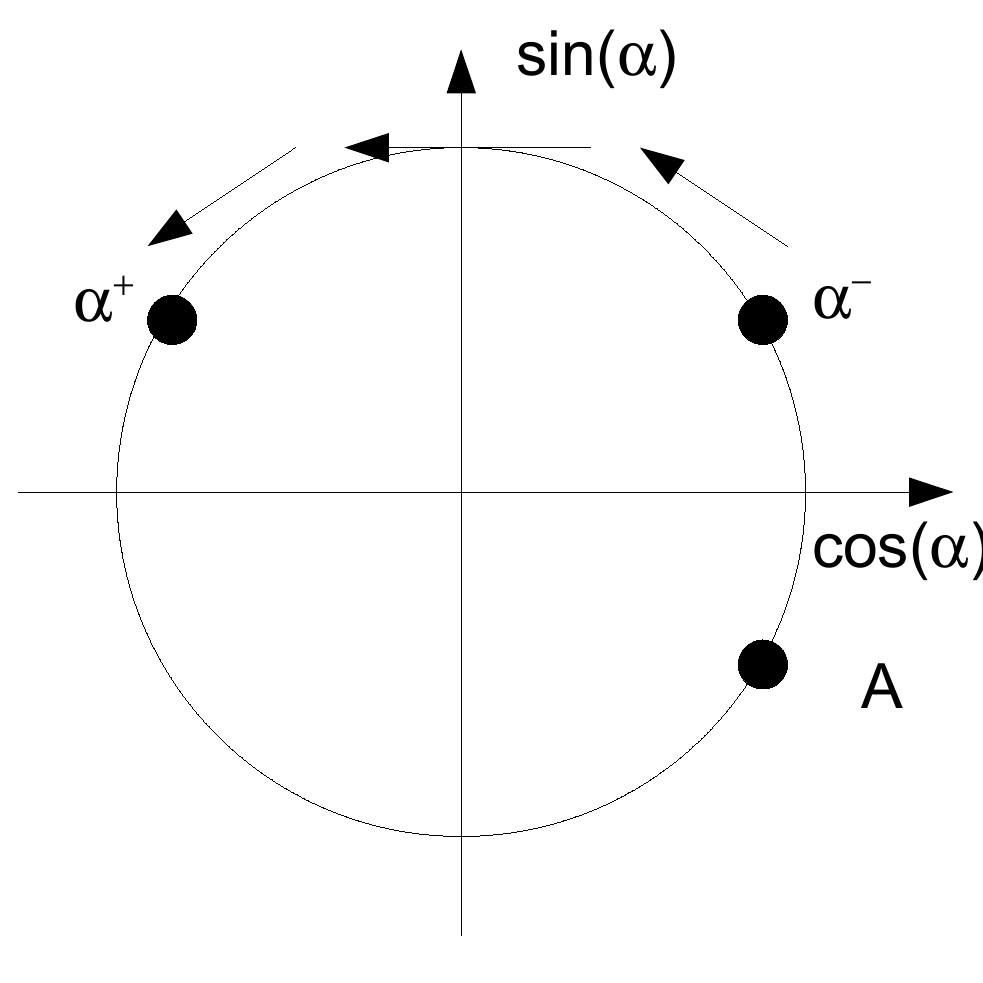}}
\end{center}
\caption{Figure showing the evolution of $e^{i\alpha'}$ during two absorption processes involving unexcited defects I and II.}
\label{fig:alpha2}
\end{figure}

As in the emission process $\alpha'$ varies with time during absorption. At the time when \begin{equation}\sqrt{1-uu^{*}}=\mathrm{sin}(A)\, ,\end{equation} the other branch of the squareroot in the definition of $\cos(\alpha')$ is taken, so $\alpha'$ changes from $-A$ in the far past to $A \pm \pi$ in the far future. Figure \ref{fig:alpha2} shows the evolution of $e^{i\alpha'}$ for the two absorption processes, figure \ref{fig:alpha2A} shows the evolution of $e^{i\alpha'}$ when unexcited defect I absorbs a soliton and figure \ref{fig:alpha2B} for unexcited defect II similarly.

Tables \ref{fig:absorption2} and \ref{fig:absorption3} shows that there are two unexcited defects that absorb the same soliton. The two unexcited defects $U_{A}$ and $U_{B}$ have the same energy and momentum but different charge $Q = -\frac{4\pi}{3}$ and $Q = \frac{2\pi}{3}$ respectively and absorb the soliton with $E_{sol}=5\sqrt{\beta}$, $P_{sol}=3\sqrt{\beta}$ and $Q_{sol}=\frac{2\pi}{3}$. With comparison with the emission tables \ref{fig:emission1} and \ref{fig:emission2}, it is found that defect $E_{A}$ emits a soliton to decay to $U_{A}$ and defect $U_{A}$ absorbs the same soliton to become $E_{A}$. 
\begin{table}[!!h]
\begin{center}
\begin{tabular}{|ccccc|}
\hline
$U_{A}$ && $\rightarrow$& &$ E_{A}$ \\
&&&&\\
$(  A =\frac{2 \pi}{3} \ , \ \delta = 2)$ &&Absorbed Soliton&&$ (  A' = \frac{\pi}{3}  , \  \delta' = 2)$ \\
$\alpha = -A $&&$(a=-A+\pi,e^{\theta}=\delta)$&&$ \alpha = A - \pi $\\
$E = -\frac{5\sqrt{\beta}}{2}$ &&$E_{sol}=5\sqrt{\beta}$&& $E' = -E$\\
$P = -\frac{3\sqrt{\beta}}{2}$ &&$P_{sol}=3\sqrt{\beta}$&& $P' = -P$\\ 
$Q = -\frac{4\pi}{3}$ &&$Q_{sol}=\frac{2\pi}{3}$&&$ Q' = -\frac{2\pi}{3}$ \\
\hline
\end{tabular}
\end{center}
\caption{Table showing unexcited defect $U_{A}$ absorbing a soliton with charge $Q_{sol}=\frac{2\pi}{3}$.}
\label{fig:absorption2}
\end{table}

\begin{table}[!!h]
\begin{center}
\begin{tabular}{|ccccc|}
\hline
$U_{B}$ && $\rightarrow$ && $E_{B}$ \\
&&&&\\
$(   A =-\frac{\pi}{3} \ ,\  \delta = -2   )$ &&Emitted Soliton&&$ ( A' = -\frac{2\pi}{3} \ , \ \delta' = -2   )$ \\
$\alpha = -A $&&$(-A,-\delta)$&&$ \alpha = A + \pi $\\
$E =-\frac{5\sqrt{\beta}}{2}$ &&$E_{sol}=5\sqrt{\beta}$&& $E' = -E$\\
$P = -\frac{3\sqrt{\beta}}{2}$ &&$P_{sol}=3\sqrt{\beta}$&& $P' = -P$\\ 
$Q = \frac{2\pi}{3}$ &&$Q_{sol}=\frac{2\pi}{3}$&&$ Q' = \frac{4\pi}{3} $\\
\hline
\end{tabular}
\end{center}
\caption{Table showing unexcited defect $U_{B}$ absorbing a soliton with charge $Q_{sol}=\frac{2\pi}{3}$.}
\label{fig:absorption3}
\end{table}

In this analysis on defects absorbing and emitting solitons we have only considered right moving soliton solutions, or at least we have only analysed them as such. In fact when solving the defect conditions no restrictions on $\theta$ were used so the solutions describe both right and left moving solitons. We have interpreted the results as solutions involving right moving solitons, but similarly there are solutions when left moving solitons are emitted or absorbed. As is the case that the sign of the defect energy determines whether a soliton can be absorbed or emitted, the sign of the defect momentum determines whether the soliton is left or right moving.

\begin{figure}[!htp]
\begin{center}
\subfigure[$E_{-} \ P_{-}$]{\label{fig:defectabsorb1}\includegraphics[width=0.23\textwidth]{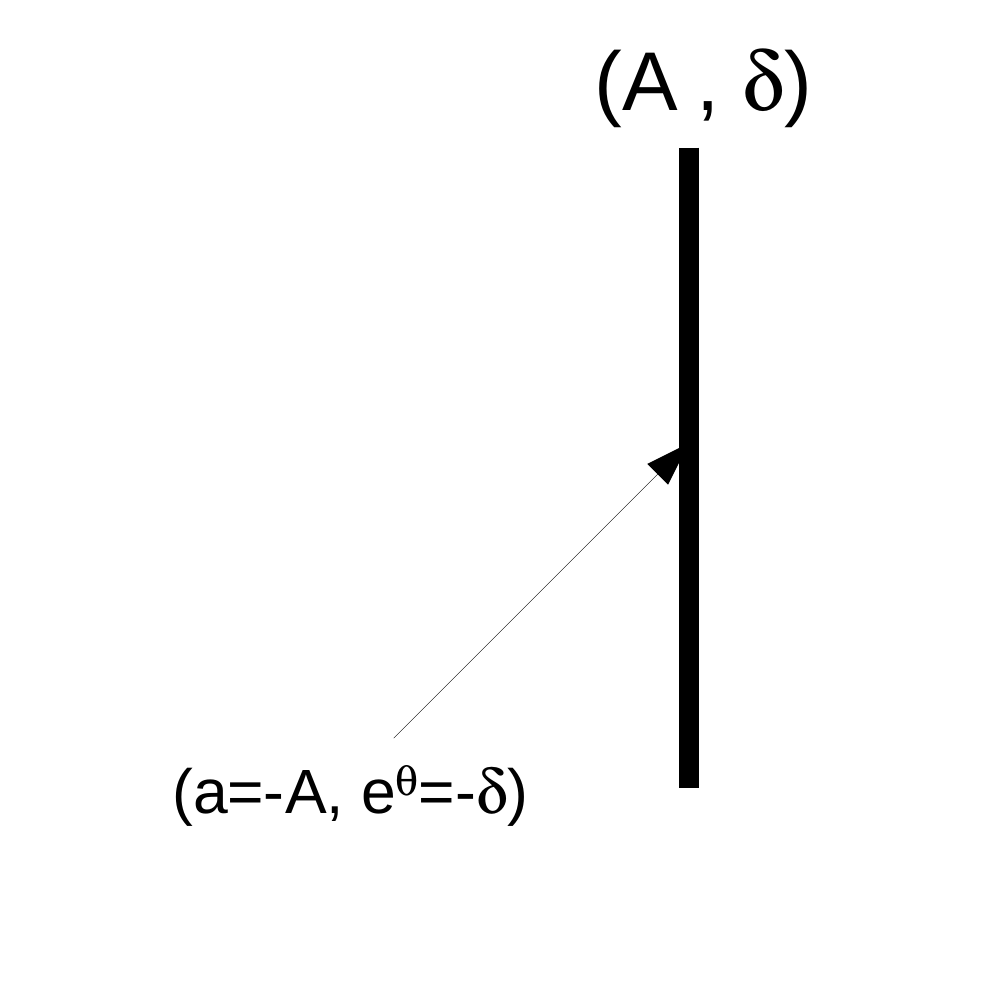}}
\subfigure[$E_{-} \ P_{+}$]{\label{fig:defectabsorb2}\includegraphics[width=0.23\textwidth]{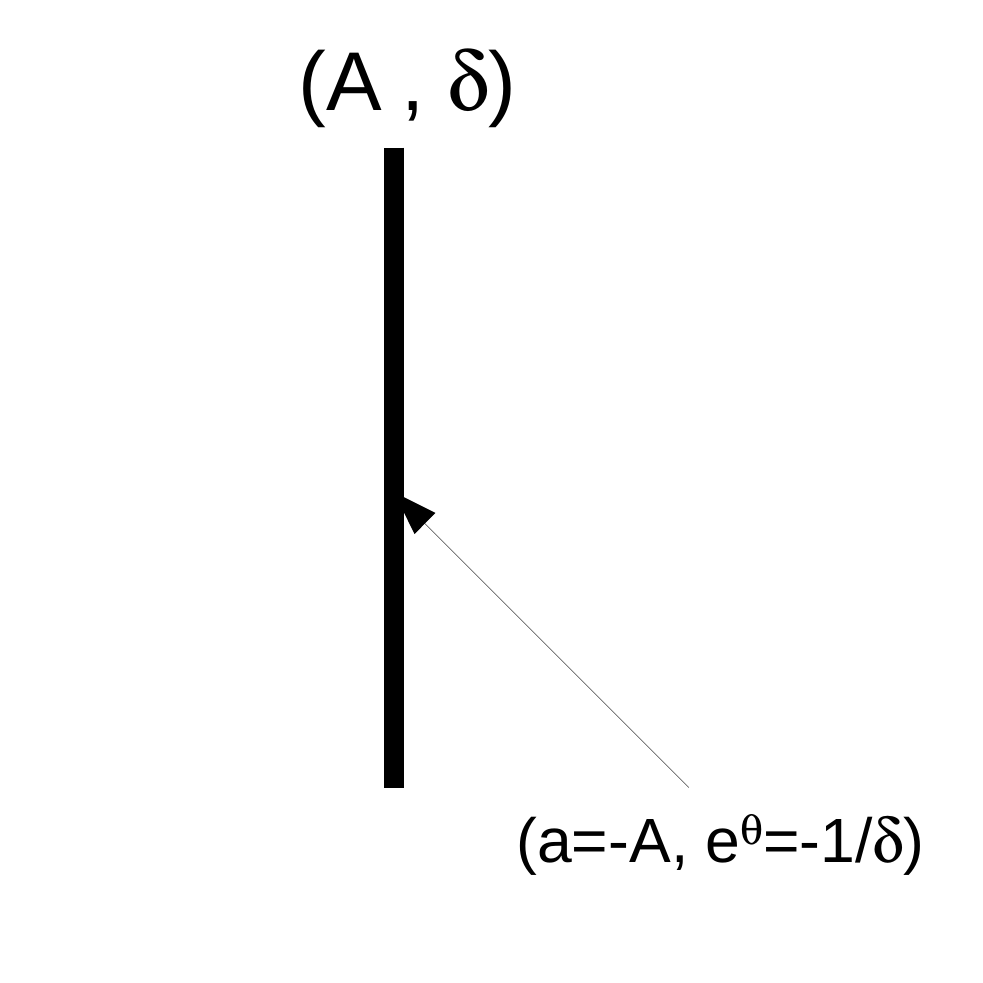}}
\subfigure[$E_{+} \ P_{-}$]{\label{fig:defectemit1}\includegraphics[width=0.23\textwidth]{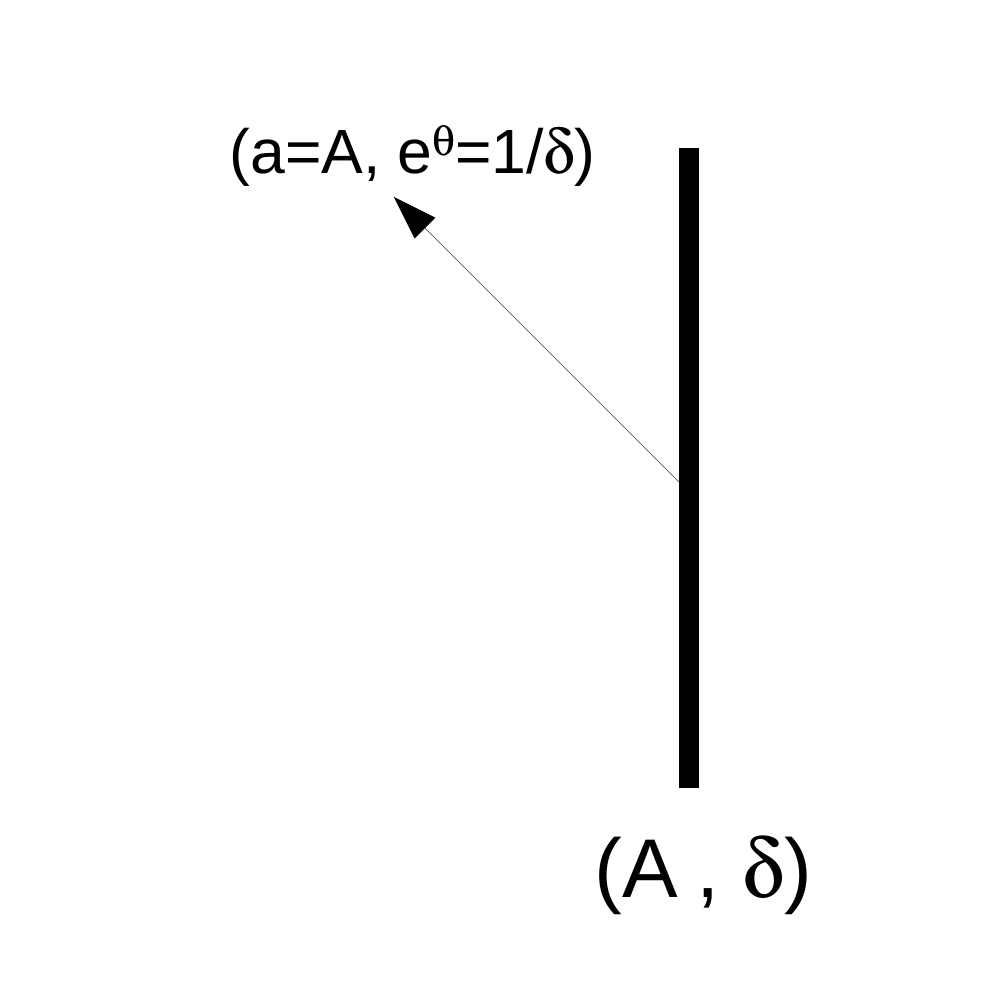}}
\subfigure[$E_{+} \ P_{+}$]{\label{fig:defectemit2}\includegraphics[width=0.23\textwidth]{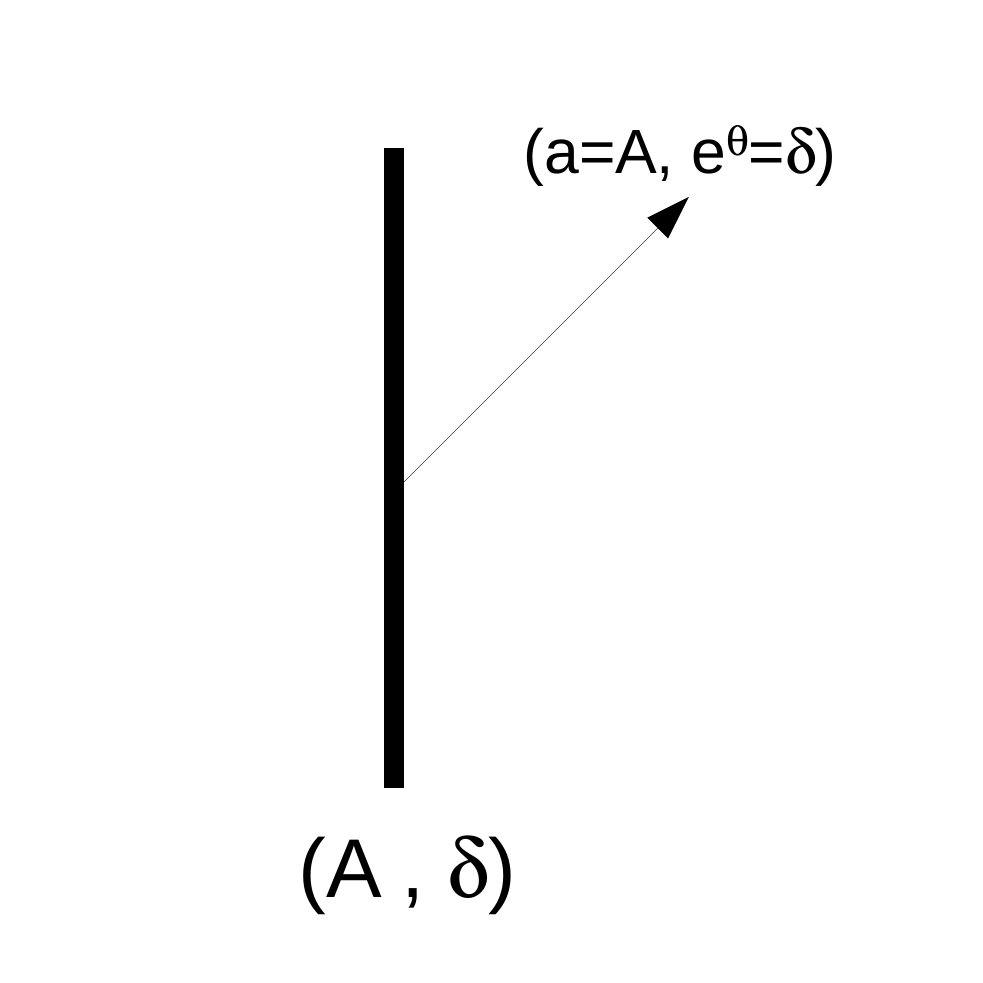}}
\end{center}
\caption{$-\frac{\pi}{2} < A < \frac{\pi}{2}$, \ \ \ (a) \ $\delta < -1$, \ \ \  (b) \ $-1 < \delta < 0$, \ \ \  (c) \ $0 < \delta < 1$, \ \ \ (d) \  $\delta > 1$.}
\label{fig:defectabsorbemit}
\end{figure}

Figure \ref{fig:defectabsorbemit}  shows explicitly that the energy and momentum of the defect determines whether it absorbs or emits a soliton and whether the soliton is right or left moving. The defect considered has charge parameter in the range $-\frac{\pi}{2} < A < \frac{\pi}{2}$ then for the four different regions of $\delta$ the soliton interaction is illustrated using $\mathrm{cos}(\alpha)=\mathrm{cos}(A)$ in the far past. In figure \ref{fig:defectabsorbemit}(a)  $\delta < - 1$ so the energy and momentum of the defect are negative, by considering only that the total energy and momentum should be conserved, it is deduced that this defect can absorb a right moving soliton. Similarly in figure \ref{fig:defectabsorbemit}(b) with  $-1 < \delta < 0$ the energy is negative and momentum is positive and the defect can absorb a left moving soliton. In figure \ref{fig:defectabsorbemit}(c) $0 < \delta < 1$  and the defect has positive energy and negative momentum so the defect can emit a left moving soliton. Finally if $\delta > 1$ as in figure \ref{fig:defectabsorbemit}(d) then both the defect energy and momentum are positive and the defect can emit a right moving soliton. The annotation in all the figures shows the condition between the soliton parameters and defect parameters for the defect conditions to be satisfied for the particular process illustrated.

 This concludes our analysis on the specific soliton solutions that different defects can emit and absorb, but what happens to the solitons which are not absorbed? 

\subsubsection{Soliton-defect scattering}

We now consider the situation where the field configuration is a one-soliton solution on each side of the defect
\begin{eqnarray}\label{eq:soldefectscat}
u &=& \frac{\mathrm{cos}(a)e^{2i\sqrt{\beta}\mathrm{sin}(a)(\mathrm{cosh}(\theta)\ t-\mathrm{sinh}(\theta)\ x)}}{\mathrm{cosh}(2\sqrt{\beta}\mathrm{cos}(a)(t\ \mathrm{sinh}(\theta)-x\ \mathrm{cosh}(\theta)))}\, ,\nonumber\\
w &=& e^{i\phi} \frac{e^{2i\sqrt{\beta}\mathrm{sin}(a)(\mathrm{cosh}(\theta)\ (t-\Delta t)-\mathrm{sinh}(\theta)\ x)}}{\mathrm{cosh}(2\sqrt{\beta}\mathrm{cos}(a)((t-\Delta t)\ \mathrm{sinh}(\theta)-x\ \mathrm{cosh}(\theta)))} \, ,
\end{eqnarray}
this is to model the process of a one-soliton scattering with the defect. The soliton solutions are not the most general but are chosen so that energy, momentum and charge are automatically conserved, i.e. the charge parameter, $a$ and rapidity, $\theta$ are taken to be the same left and right. It will be interpreted that the solitons are right moving $\theta > 0$ but again the following calculations hold for any $\theta$, note that we are allowing the outgoing soliton $w$ to differ from the incoming soliton $u$ by a time-delay, $\Delta t$ and a phase shift, $e^{i\phi}$. To calculate how a soliton is affected by travelling through a defect, the defect conditions have to be solved in this situation. The expressions for the defect conditions become extremely complicated and are solved by expanding the individual terms as a power series in the exponentiated time, $e^{2\sqrt{\beta}\mathrm{cos}(a)\mathrm{sinh}(\theta)t}$. The terms of the power series can then be solved for the time-delay 
\begin{eqnarray}\label{eq:timedelay}
\Delta t &=& \frac{1}{2\sqrt{\beta}\mathrm{cos}(a)\mathrm{sinh}(\theta)} \mathrm{ln}\left|\frac{\mathrm{sinh}\left( \frac{\theta - \chi}{2} + i \frac{a -A}{2}\right)}{\mathrm{cosh}\left(\frac{\theta - \chi}{2} +i \frac{a +A}{2}\right)}\right| \, ,
\end{eqnarray}
and phase-factor
\begin{eqnarray}\label{eq:phasefactor}
e^{i\phi} &=& e^{2\sqrt{\beta}\mathrm{sinh}(\theta+ia)\Delta t} \frac{\delta + e^{\theta}e^{ia}e^{iA}}{e^{\theta}e^{ia}-\delta e^{iA}} \, ,
\end{eqnarray}
where $e^{\chi} = \delta$ and $\alpha = -A$ at $t \rightarrow -\infty$. It is noted that this time-delay is very similar to the time-delay experienced by a soliton in soliton-soliton scattering. If the parameters are matched between the defect and one of the solitons then the time-delay experienced by a soliton scattering through the defect is exactly half of the time-delay experienced when scattering through the soliton (\ref{eq:solsolTD}). We will return to this observation later in the paper. It can be checked that the poles and zeros that appear in the logarithm in the time-delay are the exact conditions for soliton absorption and emission. This is expected as absorption and emission are different limits of the soliton-defect scattering process.

More complicated scenarios can also be analysed, for example two incoming solitons and one outgoing soliton but essentially it is just a combination of soliton scattering and a soliton being absorbed and the behaviour can be calculated from the building blocks of emission, absorption and scattering presented here.

\subsubsection{Particle-defect scattering}

We will now consider what happens to a particle travelling towards a defect. To calculate the particle-defect reflection and transmission factors one has to solve the linearised defect conditions
\begin{eqnarray}\label{eq:linearDC}
0&=& -\left(\frac{\partial u_{\epsilon}}{\partial t}-\frac{\partial u_{\epsilon}}{\partial x}\right) + \left(\frac{\partial w_{\epsilon}}{\partial t}-\frac{\partial w_{\epsilon}}{\partial x}\right)e^{i\alpha} -2\sqrt{\beta}\delta(w_{\epsilon}+u_{\epsilon} e^{i\alpha})\, ,\nonumber\\
0&=& -\left(\frac{\partial u_{\epsilon}}{\partial t}+\frac{\partial u_{\epsilon}}{\partial x}\right)e^{i\alpha} - \left(\frac{\partial w_{\epsilon}}{\partial t}+\frac{\partial w_{\epsilon}}{\partial x}\right) + \frac{2\sqrt{\beta}}{\delta}(u_{\epsilon}-w_{\epsilon} e^{i\alpha})\ ,
\end{eqnarray}
which result from substituting $u = \epsilon\  u_{\epsilon},\ \ w =  \epsilon\  w_{\epsilon}$ into the defect conditions and looking at the terms linear in $\epsilon$. 
The ansatz to which these are solved 
\begin{eqnarray}
u_{\epsilon} &=& u_{particle}(\theta) + R\ u_{particle}(-\theta)\, ,\nonumber\\
w_{\epsilon} &=& T\ u_{particle}(\theta)\, ,
\end{eqnarray}
is a left moving particle been reflected and/or transmitted by the defect, where the particle is described by $u_{particle}$ (\ref{eq:particle}). Substituting the ansatz into the linearised defect conditions (\ref{eq:linearDC}) shows that there is no reflection and the transmission factor is
\begin{eqnarray}
T_{particle/defect} &=& \frac{e^{\theta} - i\delta e^{i\alpha}}{e^{i\alpha} e^{\theta} + i\delta} \, .
\end{eqnarray}
This transmission factor can also be derived by substituting the particle limit ($a = \frac{\pi}{2}$) into the soliton-defect phase factor. In this limit the time-delay vanishes.

\subsubsection{Soliton equal to two defects?}

It was stated earlier that the soliton-defect time-delay (\ref{eq:timedelay}) is half the soliton-soliton time-delay (\ref{eq:solsolTD}) if one matches the relevant defect and soliton parameters, we will now demonstrate this more explicitly using the defects $E_A,E_B,U_A,U_B$ considered above.

\begin{figure}[!htp]
\begin{center}
\subfigure[]{\label{fig:EAEB}\includegraphics[width=0.22\textwidth]{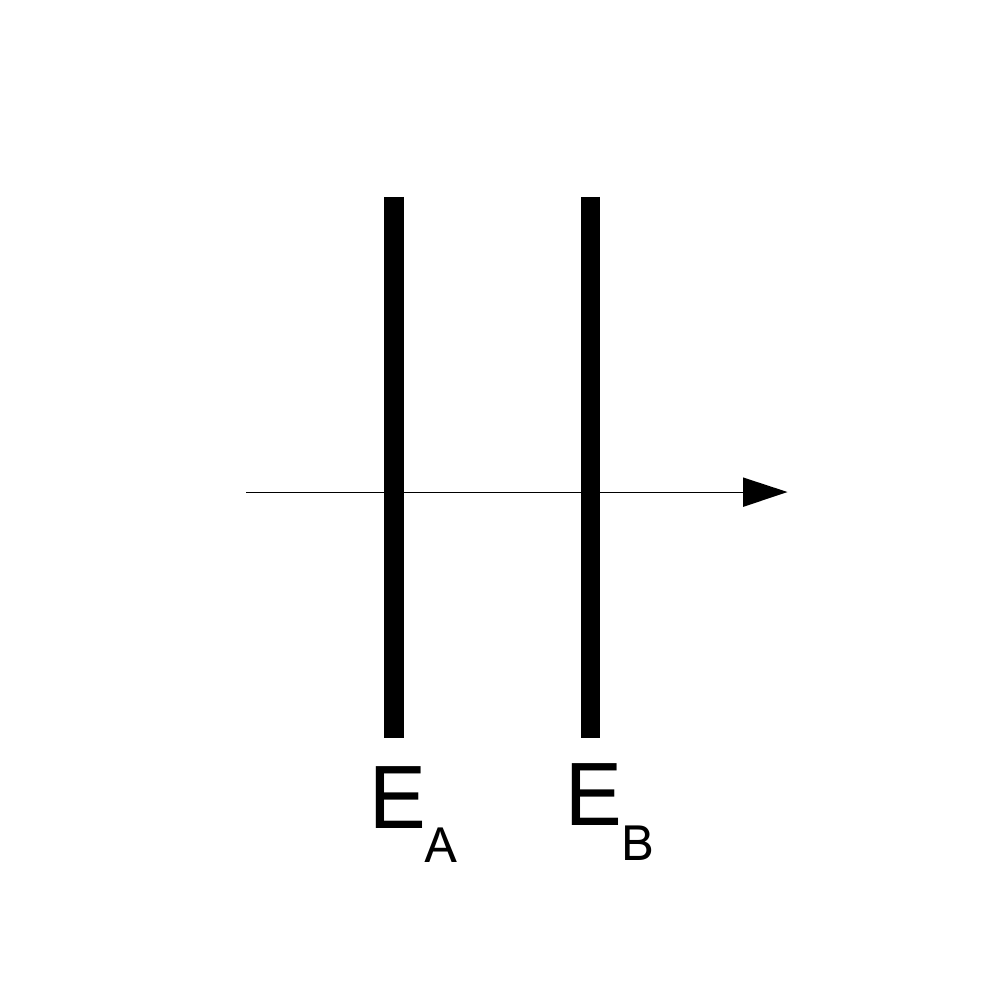}}
\subfigure[]{\label{fig:EAUB}\includegraphics[width=0.22\textwidth]{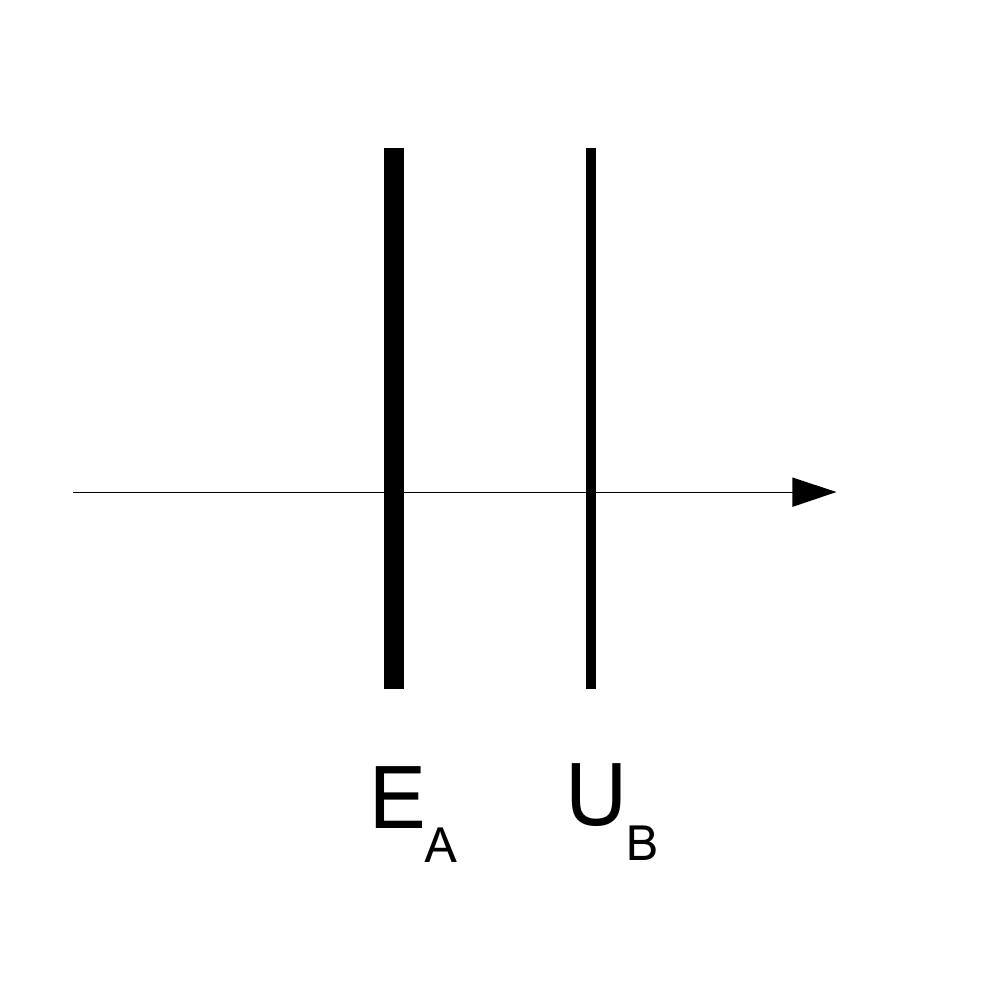}}
\subfigure[]{\label{fig:EAEBUBsol}\includegraphics[width=0.47\textwidth]{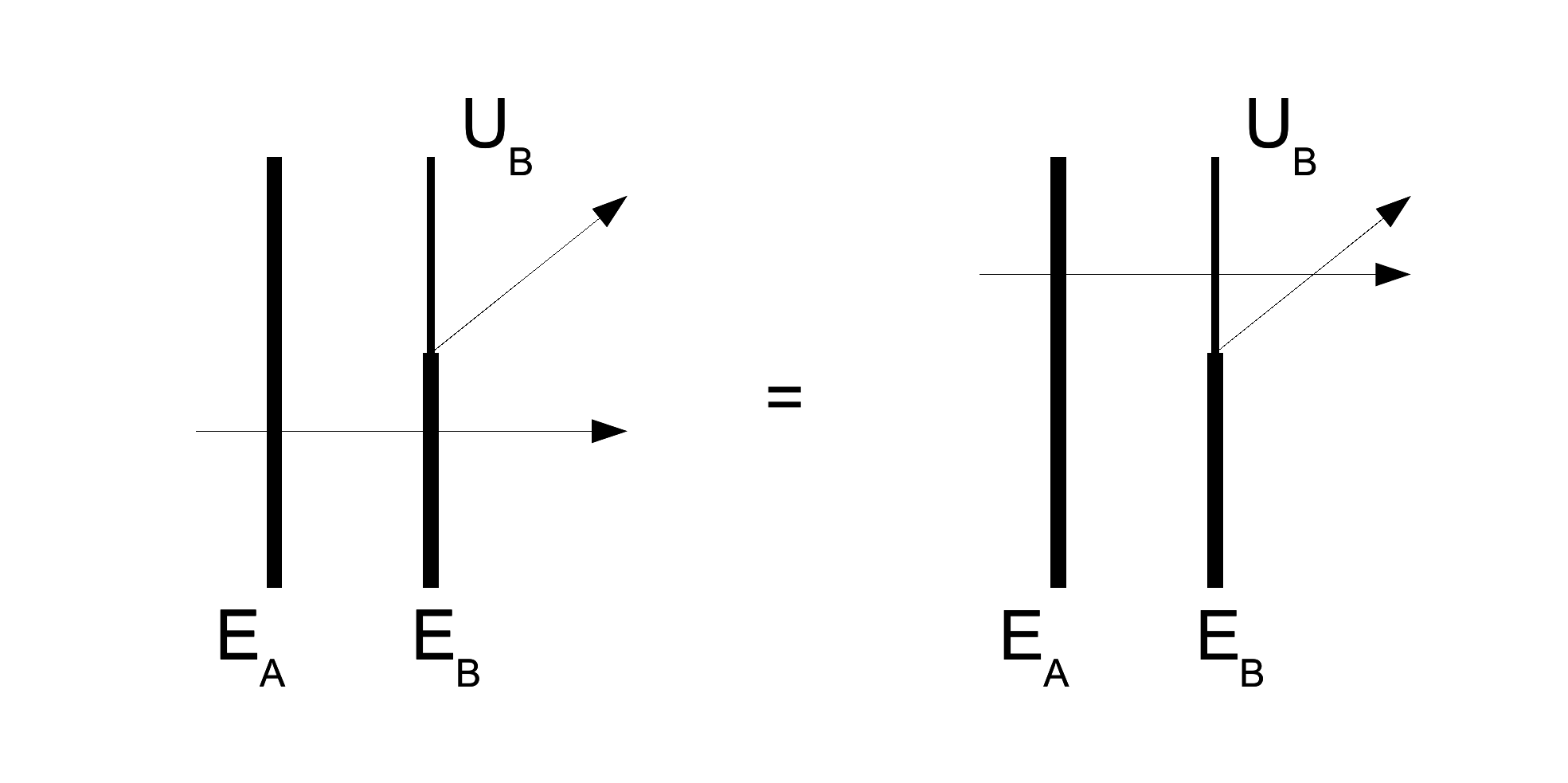}}     
\end{center}
\caption{(a) Soliton scattering through two excited defects. (b) Soliton scattering through an excited and an unexcited defect. (c) Soliton scattering through a defect and decaying defect.}
\label{fig:sol2def}
\end{figure}

It can be checked that if a soliton or particle scatters through defects $E_{A}$ and $E_{B}$ then the combined time-delay and phase factor are the same than if a soliton or particle scatters through a soliton with charge parameter $a = \frac{\pi}{3}$ and rapidity $e^{\theta} = 2$.
\begin{equation}\label{eq:2defsol} \Delta_{t}(E_{A}) + \Delta_{t}(E_{B}) = \Delta_{t}(\mathrm{sol}) \ \ \ \ \ e^{i\phi(E_{A})} e^{i\phi(E_{B})}= e^{i\phi(\mathrm{sol})}   \end{equation}
The summed total of the two defects energy, momentum and charge exactly matches the energy, momentum and charge of the soliton. These observations suggest that the integrable defects could be the fundamental objects of the theory, with two defects combined to give the solitons of the theory, this statement needs further investigation which is beyond the scope of our study here. 

If two defects are placed together whose quantities sum to zero (Figure \ref{fig:EAUB}), for example defects $E_{A}$ and $U_{B}$, then the scattering soliton or particle experiences no time-delay or phase shift. 
\begin{equation}\label{eq:2defnull} \Delta_{t}(E_{A}) + \Delta_{t}(U_{B}) = 0 \ \ \ \ \  e^{i\phi(E_{A})} e^{i\phi(U_{B})}= 1   \end{equation}
Figure \ref{fig:EAEBUBsol} shows the decay of defect $E_{B}$ when placed next to defect $E_{A}$, the integrability of the model ensures that the scattering of a soliton or particle through such a configuration should not be affected by the time that the scattering takes place. Therefore scattering through defects $E_{A}$ and $E_{B}$ should be the same as scattering through defects $E_{A}$ and $U_{B}$ and the emitted soliton, using the two previous results (\ref{eq:2defsol}, \ref{eq:2defnull}) this can be checked to be the case.

\section{Complex sine-Gordon with dressed boundary}
In this section the integrable defect will be used to create a new CSG boundary theory.
The general idea that a wider class of boundary conditions can be generated by placing a defect in front of the existing boundary has been discussed in \cite{Bowcock:2006} and used 
in the context of sinh-Gordon theory in \cite{Bajnok:2007jg} . Previously studied complex sine-Gordon boundary conditions  \cite{Bowcock:2002vz,Bowcock:2006hj} were of the form 
\begin{equation}
\label{eq:oldbc}
\partial_{x}u = -Cu\sqrt{1-uu^{*}} \, ,
\end{equation} 
which depend on only one parameter. Dressed boundary conditions will inherit two parameters from the defect, and this suggests that they form a more general class of boundary conditions.

\subsection{Constructing the dressed boundary theory}
\begin{figure}[!h]
\begin{center}
\includegraphics[width=0.75\textwidth]{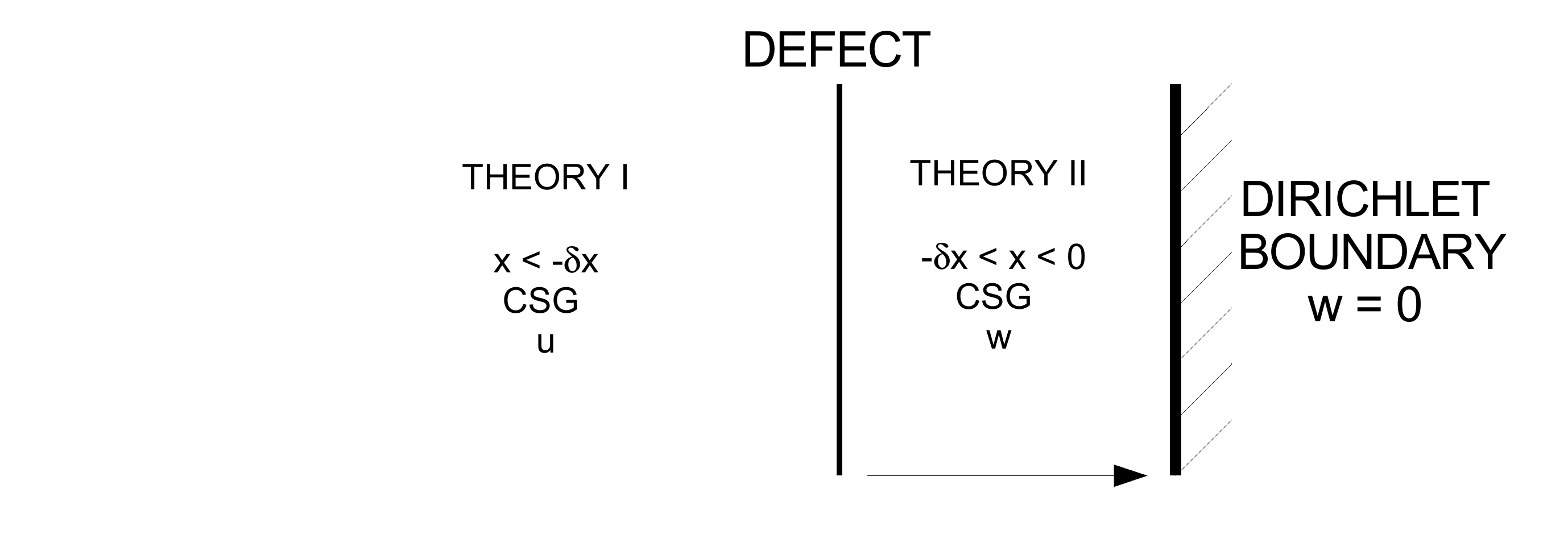}
\end{center}
\caption{Dressed boundary model set-up.}
\label{Fig:dressedboundary}
\end{figure}

We start with a Dirichlet boundary with the boundary condition $w=0$ and place an integrable defect in front of it as shown in figure \ref{Fig:dressedboundary}.
The integrability of the model allows us to move the defect up to the boundary to create the dressed boundary. To formulate the dressed boundary conditions the Dirichlet boundary condition $w=0, \ \partial_{t}w=0$ are substituted into the defect conditions (\ref{eq:DefectCond}). Then $\partial_{x} w$ is eliminated by combining the defect conditions to give the dressed boundary conditions in terms of the complex fields $u,\ u^{*}$
\begin{eqnarray}\label{eq:dressedBC1}
\partial_{x} u &=& -\partial_{t}u \ i\  \mathrm{tan}(\alpha') + \frac{\sqrt{\beta}}{\mathrm{cos}(\alpha')}\left(\delta+\frac{1}{\delta}\right)u\sqrt{1-uu^{*}} \, ,\nonumber\\
\partial_{x} u^{*} &=& \partial_{t}u^{*} \ i\  \mathrm{tan}(\alpha') + \frac{\sqrt{\beta}}{\mathrm{cos}(\alpha')}\left(\delta+\frac{1}{\delta}\right)u^{*}\sqrt{1-uu^{*}} \, .
\end{eqnarray}
where \begin{eqnarray} \alpha' &=& \mathrm{arcsin}\left(-\frac{\mathrm{sin}(A)}{\sqrt{1-uu^{*}}}\right)\, . \end{eqnarray} 
These boundary conditions depend on the two parameters $\delta, A$ that appear in the B\"{a}cklund transformation. To find the expression for the  dressed boundary Lagrangian we vary the action to produce the boundary Euler-Lagrange equations which gives the normal equation of motion in the bulk region $x<0$ and the boundary conditions at $x=0$
\begin{eqnarray}\label{eq:E-L}
\frac{\partial_{x} u}{1-uu^{*}} &=& \frac{ \partial \mathcal{L}_{B}}{\partial u^{*}} -\partial_{t} \left(\frac{\partial \mathcal{L}_{DB}}{\partial \partial_{t}u^{*}} \right)\, ,\nonumber \\
\frac{\partial_{x} u^{*}}{1-uu^{*}} &=& \frac{ \partial \mathcal{L}_{B}}{\partial u}-\partial_{t} \left(\frac{\partial \mathcal{L}_{DB}}{\partial \partial_{t}u} \right) \, .
\end{eqnarray}
Using the defect theory as a guide we assume that the dressed boundary Lagrangian has the form
\begin{eqnarray}\label{eq:dressedboundarylagrangianA}
 \mathcal{L}_{dCSG}& =& \int^{0}_{-\infty} dx\ \  \frac{ \partial_{t} u \partial_{t}u^{*} -\partial_{x}u \partial_{x} u^{*} }{1-   u u^{*}} - 4\beta uu^{*}+ \left.\left[A_{1} \partial_{t}u + A_{2} \partial_{t}u^{*}  -\mathcal{L}_{db}\right]\right|_{x=0} \, ,
\end{eqnarray}
with the standard bulk Lagrangian and a boundary piece made up a boundary potential term and terms in linear in the time derivative of the field. The unknown quantities $A_{i}$ and $\mathcal{L}_{b}$ can be found, by comparing the dressed boundary conditions (\ref{eq:dressedBC1}) and the Euler-Lagrange equations (\ref{eq:E-L}) where $\mathcal{L}_{DB} = A_{1} \partial_{t}u + A_{2} \partial_{t}u^{*}  -\mathcal{L}_{db}$, to be 

\begin{eqnarray}
\mathcal{L}_{db} &=&  2\sqrt{\beta}\left(\delta + \frac{1}{\delta}\right) \mathrm{cos}(\alpha')\sqrt{ 1 -uu^{*}}\, ,\nonumber\\
A_{1} &=& -\frac{i}{u}\alpha' \, , \ \ \ \  A_{2} = \frac{i}{u^{*}}\alpha'\,. \nonumber\\
\end{eqnarray}
As with the defect, the boundary energy can be read directly from the Lagrangian 
\begin{eqnarray}\label{eq:boundaryenergy}
E_{dbound} &=& \int^{0}_{-\infty} dx \ \ \frac{ \partial_{t}u \partial_{t} u^{*} + \partial_{x}u\partial_{x} u^{*}}{1-uu^{*}} + 4\beta uu^{*} + \left.\left[ 2\sqrt{\beta}\left(\delta + \frac{1}{\delta}\right) \mathrm{cos}(\alpha')\sqrt{ 1 -uu^{*}} \right]\right|_{x=0}
\end{eqnarray}
 and its conservation can be easily checked. The boundary charge can also be calculated
\begin{eqnarray}\label{eq:boundarycharge}
Q_{dbound} &=& i\int^{0}_{-\infty} dx \ \ \frac{u\partial_{t}u^{*} -u^{*}\partial_{t}u}{1-uu^{*}}  +\left.\left[2\alpha'  \right]\right|_{x=0}\, ,
\end{eqnarray}
using the dressed boundary conditions to specify a boundary term needed to be added to the bulk charge to make it conserved. In the dressed boundary theory there is not a conserved dressed boundary momentum as is usual for theories with a boundary. The lack of conserved momentum does not stop the theory being classically integrable. The conservation of all the higher spin energy-like charges are assumed to follow from the construction of the boundary from an integrable boundary condition and integrable defect.

There is a connection between the dressed boundary quantities and their defect counterparts.  The dressed boundary Lagrangian and quantities are retrieved from the defect ones by substituting in the Dirichlet conditions $w=0, \ \  \partial_{t}w=0$. This shows that as well as by first finding the dressed boundary conditions, the dressed boundary theory can be constructed directly in the Lagrangian picture. In recent work Bajnok and Simon have demonstrated that this works for the sinh-Gordon theory \cite{Bajnok:2007jg}, although it is not totally obvious how to write down the Lagrangian for the theory with Dirichlet boundary. We also note that  we can recover the previously known boundary condition (\ref{eq:oldbc}) as the  $A=0$ limit of the dressed boundary conditions.

\subsection{Soliton solutions in the dressed boundary theory}
In this section we will look at the different soliton solutions that solve the field equations, the equations of motion and dressed boundary conditions. We will examine the vacuum of the theory before investigating whether the boundary can absorb or emit solitons, look for boundary bound states and examine how solitons and particles scatter from the boundary.

\subsubsection{Vacuum}

The vacuum of the dressed theory is $u=0$ with energy and charge
\begin{eqnarray}
E_{vac}& =&2\sqrt{\beta}\left(\delta + \frac{1}{\delta}\right)  \mathrm{cos}(\alpha_0)\, ,\nonumber\\
Q_{vac} &=& 2\alpha_0 \, ,
\end{eqnarray}
where $\alpha_0 = \alpha'(u=0)$. As with the defect theory the energy and charge of the vacuum depend on the two parameters that appear in the B\"{a}cklund transformation and again explicitly on $\alpha_0$. This $\alpha_0$ dependence means that the dressed boundary can have different values of energy and charge for the same values of $\delta$ and $A$. As in the defect theory whether the energy of the dressed boundary is positive or negative determines its properties. 

Analogous to the defect theory, we expect a boundary with positive energy to be able to emit a soliton, while a boundary of negative energy to absorb a soliton. We will analyse these scenarios in the next section, but from now on we will label a boundary with negative energy as an unexcited boundary and a boundary with positive energy an excited boundary.

It is natural to split the excited boundaries into eight types, meaning that there are eight boundaries with the same positive energy but different charges. In fact pairs of the boundaries have the same charge, but we will list them separately as they are described by different $\delta$ and $A$. The excited boundaries listed in table \ref{exbound} have energy $E = 2\sqrt{\beta}|\left(\delta + \frac{1}{\delta}\right)\mathrm{cos}(A)|$. For the specific case where $A'=\frac{\pi}{3}$ and $\delta'=2$ then the eight boundaries have energy $E= \frac{5\sqrt{\beta}}{2}$ and charges $Q \in \left\{ \pm \frac{2\pi}{3} , \ \ \pm \frac{4\pi}{3}\right\}$.  

\begin{table}[h]
\begin{center}
\begin{tabular}{|rcl|l|l|l|ll|}
\hline
$0 <$&$ A$&$ < \frac{\pi}{2}$&$A=A'$ &$ \delta > 1$ &$\delta=\delta'$&$ Q =-2A'$&$\in \left\{ 0,-\pi \right\}$\\
$- \frac{\pi}{2} <$&$ A$&$ < 0$ &$A=-A'$&$ \delta > 1 $&$\delta=\delta'$&$ Q =2A'$&$\in \left\{ \pi,0 \right\}$ \\
$\frac{\pi}{2} <$&$ A$&$ < \pi $&$A=\pi-A'$&$ \delta < -1$ &$\delta=-\delta'$&$ Q =2(A'-\pi)$&$\in \left\{ -\pi,-2\pi \right\} $\\
$-\pi <$&$ A$&$ < -\frac{\pi}{2}$ &$A=-\pi+A'$&$ \delta < -1$ &$\delta=-\delta'$&$ Q =2(\pi-A')$&$\in \left\{ 2\pi,\pi \right\} $\\
$0 <$&$ A$&$ < \frac{\pi}{2}$ &$A=A'$&$ 0 < \delta < 1$ &$\delta=\frac{1}{\delta'}$&$ Q=-2A'$&$ \in \left\{ 0,-\pi \right\}$ \\
$- \frac{\pi}{2} <$&$ A$&$ < 0$ &$A=-A'$&$ 0 < \delta < 1 $&$\delta=\frac{1}{\delta'}$&$ Q =2A'$&$\in \left\{ \pi,0 \right\}$ \\
$\frac{\pi}{2} <$&$ A$&$ < \pi $&$A=\pi-A'$&$ -1 < \delta < 0$ &$\delta=-\frac{1}{\delta'}$&$ Q=2(A'-\pi)$&$ \in \left\{ -\pi,-2\pi \right\} $\\
$-\pi <$&$ A$&$ < -\frac{\pi}{2}$ &$A=-\pi+A'$&$ -1 < \delta < 0$ &$\delta=-\frac{1}{\delta'}$&$ Q =2(\pi-A')$&$\in \left\{ 2\pi,\pi \right\} $\\
\hline
\end{tabular}
\end{center}
\caption{Listing 8 excited boundaries with the same positive energy.}
\label{exbound}
\end{table}

 Similarly there are eight unexcited boundaries with energy $E = -2\sqrt{\beta}|\left(\delta + \frac{1}{\delta}\right)\mathrm{cos}(A)|$ displayed table \ref{unexbound}. The specific case when $A'=\frac{\pi}{3}$ and $\delta'=-2$ gives boundaries with $E=-\frac{5\sqrt{\beta}}{2}$ and charges $Q \in \left\{ \pm \frac{2\pi}{3} , \ \ \pm \frac{4\pi}{3}\right\}$.

\begin{table}[h]
\begin{center}
\begin{tabular}{|rcl|l|l|l|ll|}
\hline
$0 < $&A&$ < \frac{\pi}{2}$ &$A=A'$&$ \delta < -1$ &$\delta=\delta'$&$ Q =-2A'$&$\in \left\{ 0,-\pi \right\}$ \\
$- \frac{\pi}{2} <$& A&$ < 0$ &$A=-A'$&$ \delta < -1 $&$\delta=\delta'$&$ Q =2A'$&$\in \left\{ \pi,0 \right\}$ \\
$\frac{\pi}{2} <$& A&$ < \pi $&$A=\pi-A'$&$ \delta > 1$ &$\delta=-\delta'$&$ Q =2(A'-\pi)$&$\in \left\{ -\pi,-2\pi \right\} $\\
$-\pi <$& A&$ < -\frac{\pi}{2}$ &$A=A'-\pi$&$ \delta > 1$ &$\delta=-\delta'$&$ Q =2(\pi-A')$&$\in \left\{ 2\pi,\pi \right\} $\\
$0 < $&A&$ < \frac{\pi}{2}$ &$A=A'$&$ -1 <\delta < 0$ &$\delta=\frac{1}{\delta'}$&$ Q =-2A'$&$\in \left\{ 0,-\pi \right\}$ \\
$- \frac{\pi}{2} <$& A&$ < 0$ &$A=-A'$&$ -1 <\delta < 0 $&$\delta=\frac{1}{\delta'}$&$ Q =2A'$&$\in \left\{ \pi,0 \right\}$ \\
$\frac{\pi}{2} <$& A&$ < \pi $&$A=\pi-A'$&$ 0 < \delta < 1$ &$\delta=-\frac{1}{\delta'}$&$ Q =2(A'-\pi)$&$\in \left\{ -\pi,-2\pi \right\} $\\
$-\pi <$& A&$ < -\frac{\pi}{2}$ &$A=A'-\pi$&$ 0 < \delta < 1$ &$\delta=-\frac{1}{\delta'}$&$ Q =2(\pi-A')$&$\in \left\{ 2\pi,\pi \right\} $\\
\hline
\end{tabular}
\end{center}
\caption{Listing 8 unexcited boundaries with the same negative energy.}
\label{unexbound}
\end{table}

\subsubsection{Soliton absorption by the boundary}
To analyse whether a soliton can be absorbed by the dressed boundary, the field equations are solved for a one-soliton solution moving from left infinity towards the boundary. Using $\mathrm{cos}(\alpha') = \mathrm{cos}(A)$ at the past temporal infinity, the dressed boundary conditions are satisfied by

\begin{equation}
\label{eq:BoundaryAbs}
\begin{array}{rclcrcl}
\delta& =& e^{\theta},\ e^{-\theta}& & a& =& -A + \pi \, ,\\
\delta& =& -e^{\theta},\ -e^{-\theta}& & a& =& -A \, .\\ 
\end{array}
\end{equation} 
Since the boundary conditions can be satisfied in this set-up there are boundaries that do absorb a soliton. This is expected knowing the behaviour of the defect, which also absorbs solitons and remembering that the dressed boundary is constructed by placing a defect in front of a Dirichlet boundary. Recall that the defect could only absorb a left moving soliton if the defect rapidity was $|\delta| > 1$, but the boundary can absorb a soliton for any value of $\delta$.

\begin{figure}[!h]
\begin{center}
\subfigure[]{\label{fig:boundabsorb1}\includegraphics[width=0.25\textwidth]{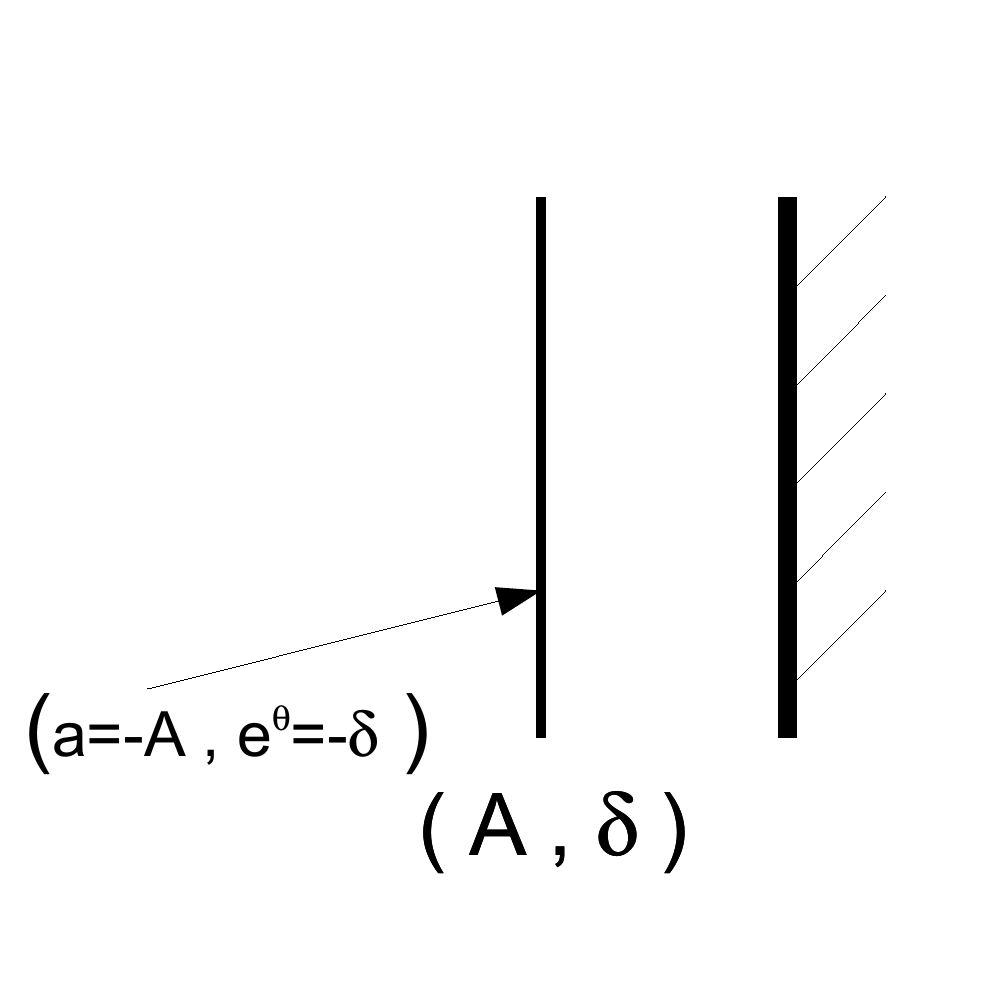}}
\hspace{0.8in}
\subfigure[]{\label{fig:boundabsorb2}\includegraphics[width=0.25\textwidth]{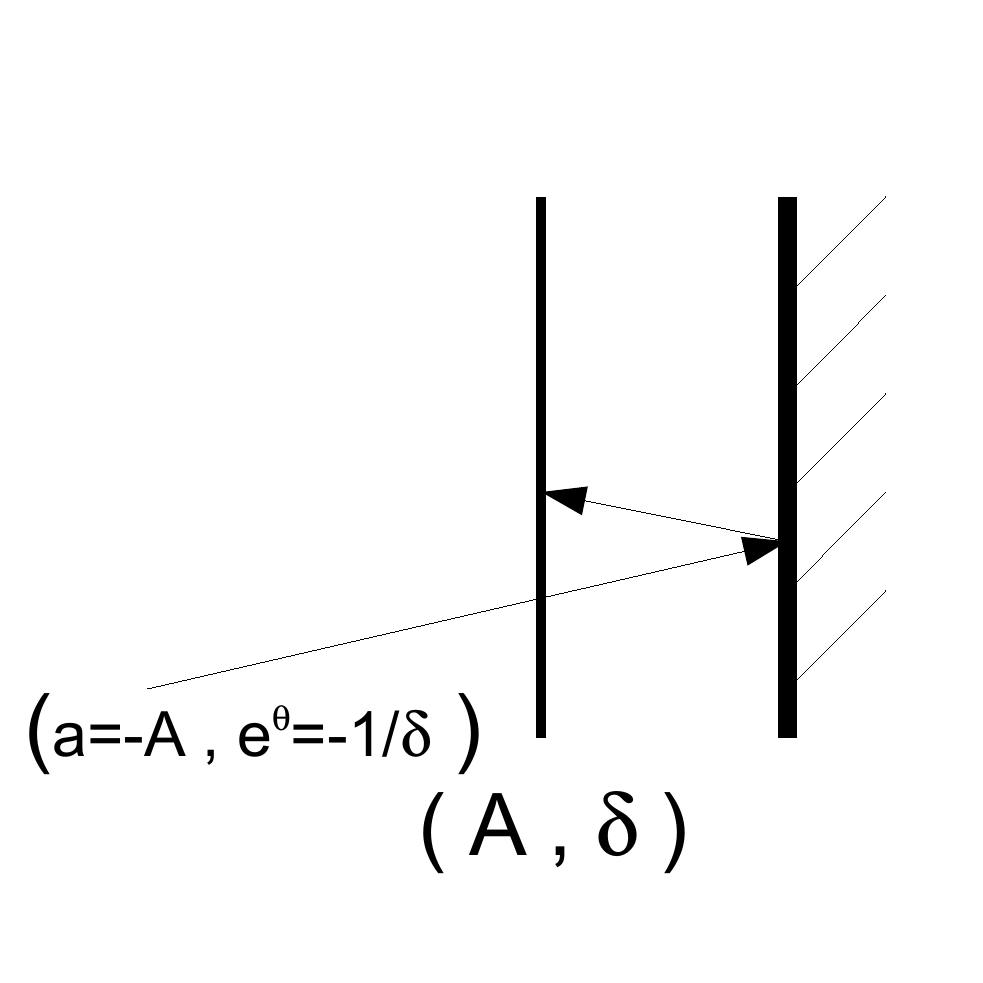}}
\end{center}
\caption{Showing the two methods of soliton absorption by the dressed boundary.}
\label{fig:boundabsorb}

\end{figure}

By thinking of the dressed boundary as a construction of its constituent defect and boundary there are two ways that a boundary can absorb a soliton. These two absorption methods are shown in figure \ref{fig:boundabsorb}.
Figure \ref{fig:boundabsorb1} shows the soliton being absorbed directly by the defect and figure \ref{fig:boundabsorb2} shows the second method of absorption where the soliton is absorbed after reflecting back from the Dirichlet boundary. This second method allows boundaries with rapidity $|\delta| < 1$ to absorb solitons, as well as the usual defect absorption with rapidity $|\delta| > 1$.

Each boundary absorbs exactly one soliton but there are four unexcited boundaries that absorb a soliton of the same energy and charge. Tables \ref{absorbEG} and \ref{absorbEG1}  shows two such boundaries absorbing a charge $Q_{sol}=\frac{2\pi}{3}$ soliton with energy $E_{sol} = 5\sqrt{\beta}$, the other two boundaries that absorb the same soliton are related by $\delta \rightarrow \frac{1}{\delta}$. In absorption I we have an unexcited boundary described by $A=\frac{2\pi}{3}$ and $\delta=2$ being excited by the absorption of said soliton and similarly in II an unexcited boundary with $A =-\frac{\pi}{3}$ and $\delta=-2$. Analogous to the defect case $\alpha'$ changes in time allowing the boundary to change energy and charge which makes this absorption process possible.

\begin{table}[h]
\begin{center}
\begin{tabular}{|ccccc|}
\hline
Unexcited Boundary I && $\rightarrow$& & Excited Boundary I \\
&&&&\\
$( A = \frac{2\pi}{3} \ , \ \delta = 2)$ &&Absorbed Soliton&&$ (  A' = \frac{\pi}{3}  , \  \delta' = 2)$ \\
$\alpha = -A $&&$(a=-A+\pi,e^{\theta}=\delta)$&&$ \alpha = A - \pi $\\
$E = -\frac{5\sqrt{\beta}}{2}$ &&$E_{sol}=5\sqrt{\beta}$&&$ E' = -E$\\

$Q = -\frac{4\pi}{3}$ &&$Q_{sol}=\frac{2\pi}{3}$&&$ Q' = -\frac{2\pi}{3} $\\
\hline
\end{tabular}
\end{center}
\caption{Showing unexcited boundary I absorbing a soliton with $E_{sol}=5\sqrt{\beta}$ and $Q_{sol}=\frac{2\pi}{3}$.}
\label{absorbEG}
\end{table}

\begin{table}[h]
\begin{center}
\begin{tabular}{|ccccc|}
\hline
Unexcited Boundary II &&$ \rightarrow$ && Excited Boundary II \\
&&&&\\
$(   A =-\frac{\pi}{3} \ ,\  \delta = -2   )$ &&Absorbed Soliton&&$ (  A' = -\frac{2\pi}{3} \ , \ \delta' = -2   )$ \\
$\alpha = -A $&&$(a=-A,e^{\theta}=-\delta)$&&$ \alpha = A + \pi $\\
$E =-\frac{5\sqrt{\beta}}{2}$ &&$E_{sol}=5\sqrt{\beta}$&&$ E' = -E$\\

$Q = \frac{2\pi}{3}$ &&$Q_{sol}=\frac{2\pi}{3}$&&$ Q' = \frac{4\pi}{3}$ \\
\hline
\end{tabular}
\end{center}
\caption{Showing unexcited boundary II absorbing a soliton with $E_{sol}=5\sqrt{\beta}$ and $Q_{sol}=\frac{2\pi}{3}$.}
\label{absorbEG1}
\end{table}

\subsubsection{Soliton emission}

\begin{figure}[!h]
\begin{center}
\subfigure[]{\label{fig:boundemit1}\includegraphics[width=0.25\textwidth]{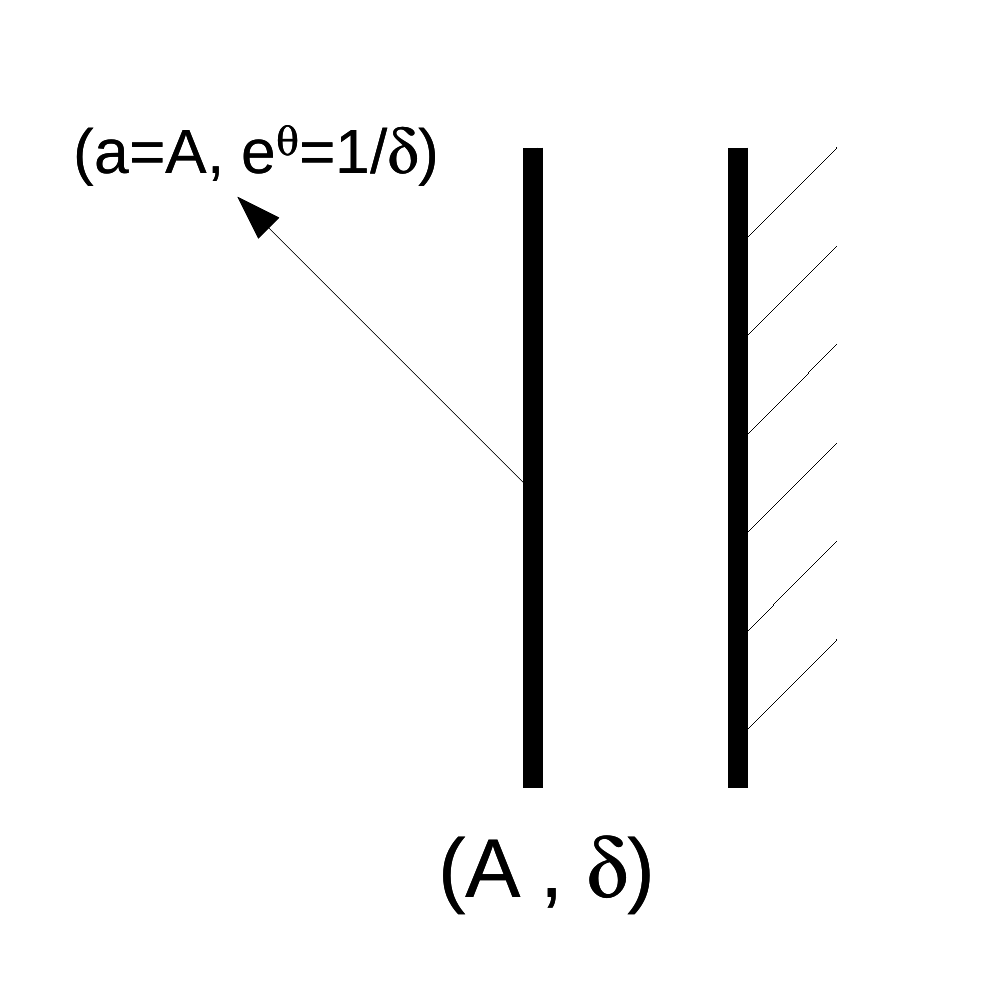}}
\hspace{0.8in}
\subfigure[]{\label{fig:boundemit2}\includegraphics[width=0.25\textwidth]{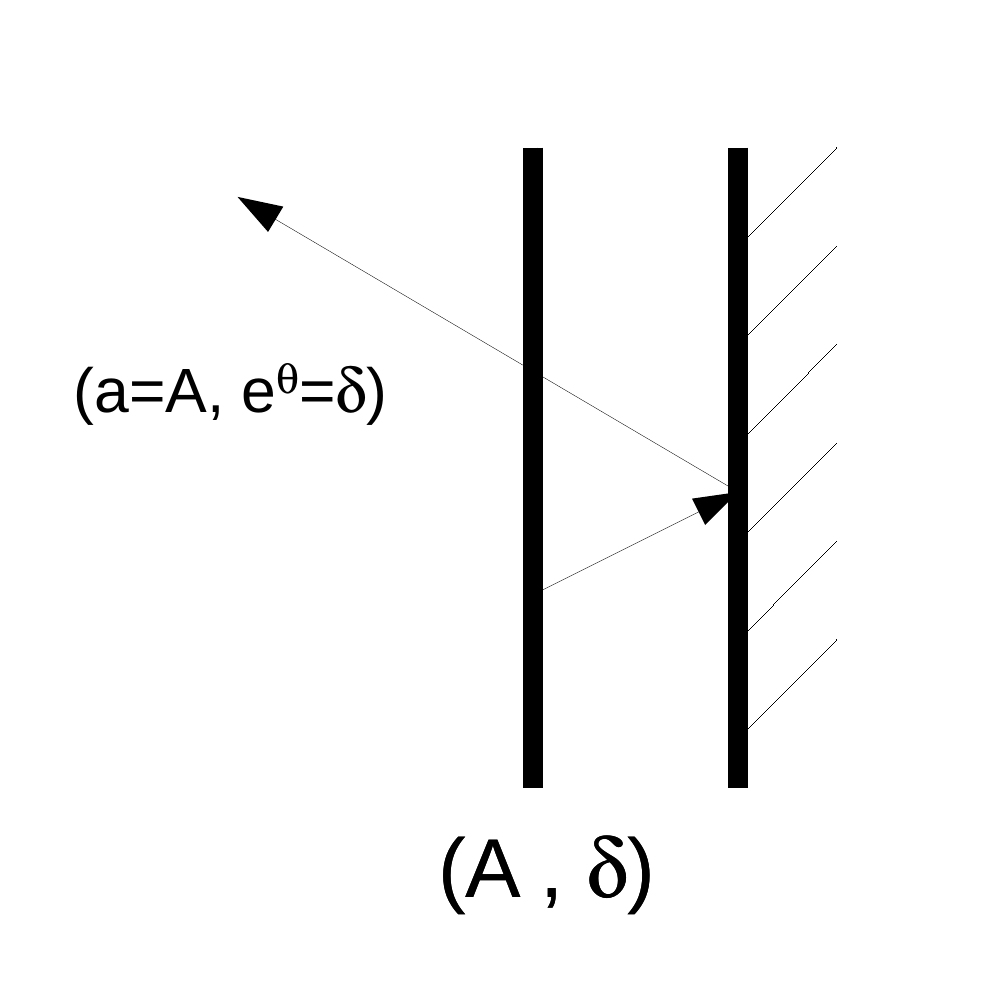}}
\end{center}
\caption{Showing the two methods of soliton emission by the dressed boundary.}
\label{fig:boundemit}
\end{figure}

To examine whether the boundary can emit a soliton the field equations are solved for a left moving soliton, once again with the choice $\mathrm{cos}(\alpha') = \mathrm{cos}(A)$ at $t \rightarrow -\infty$. The dressed boundary conditions are satisfied by

\begin{equation}
\label{eq:boundaryEm}
\begin{array}{rclcrcl}
\delta& =& e^{\theta},\ e^{-\theta}& & a& =& A \, , \\
\delta& =& -e^{\theta},\ -e^{-\theta}& & a& =& A+\pi \, .\\ 
\end{array}
\end{equation} 
As with the absorption process there are two ways in which the soliton can be emitted, either emitted directly from the defect, \ref{fig:boundemit1}, or emitted to the right before reflecting back from the Dirichlet boundary, \ref{fig:boundemit2}. These two methods of emission explain why a boundary described by any $\delta \in \mathbb{R}$ can emit a soliton.

Again each boundary emits just one type of soliton but there are four boundaries that emit a soliton with the same energy and charge. Tables \ref{emitEG} and \ref{emitEG1} show two boundaries that emit a soliton with energy $E_{sol}=5\sqrt{\beta}$ and charge $Q_{sol}=\frac{2\pi}{3} $. The two boundaries shown in the tables, excited boundary I described by $A=\frac{\pi}{3}$ and $\delta=2$ and excited boundary II with $A = -\frac{2\pi}{3}$ and $ \delta=-2$ are joined by two other boundaries related by $\delta \rightarrow \frac{1}{\delta}$ that emit the same soliton.

\begin{table}[h]
\begin{center}
\begin{tabular}{|ccccc|}
\hline
Excited Boundary I && $\rightarrow$& & Unexcited Boundary I \\
&&&&\\
$(  A =\frac{\pi}{3} \ , \ \delta = 2)$ &&Emitted Soliton&&$ (  A' = \frac{2\pi}{3} \ , \  \delta' = 2)$ \\
$\alpha = -A $&&$(a=A,e^{\theta}=\delta)$&&$ \alpha = A-\pi $\\
$E = \frac{5\sqrt{\beta}}{2}$ &&$E_{sol}=5\sqrt{\beta}$&& E' = -E\\
$Q = -\frac{2\pi}{3} $&&$Q_{sol}=\frac{2\pi}{3} $&&$ Q' = -\frac{4\pi}{3}$ \\
\hline
\end{tabular}
\end{center}
\caption{Showing excited boundary I emitting a soliton with $E_{sol}=5\sqrt{\beta}$ and $Q_{sol}=\frac{2\pi}{3}$.}
\label{emitEG}
\end{table}

\begin{table}[h]
\begin{center}
\begin{tabular}{|ccccc|}
\hline
Excited Boundary II &&$ \rightarrow$ && Unexcited Boundary II \\
&&&&\\
$( A =-\frac{2\pi}{3} \ ,\  \delta = -2   )$ &&Emitted Soliton&&$ (  A' = -\frac{\pi}{3}  \ , \ \delta' = -2   )$ \\
$\alpha = -A $&&$(a=A+\pi,-e^{\theta}=\delta)$&&$ \alpha = A+\pi $\\
$E = \frac{5\sqrt{\beta}}{2}$ &&$E_{sol}=5\sqrt{\beta}$&& E' = -E\\
$Q = \frac{4\pi}{3}$&&$ Q_{sol}=\frac{2\pi}{3}$&&$ Q' = \frac{2\pi}{3}$ \\
\hline
\end{tabular}
\end{center}
\caption{Showing excited boundary II emitting a soliton with $E_{sol}=5\sqrt{\beta}$ and $Q_{sol}=\frac{2\pi}{3}$.}
\label{emitEG1}
\end{table}

The change in charge during emission and absorption is once again symmetric around $Q = \pm \pi$, for example during the emission process described in table \ref{emitEG1} the charge of the boundary before is $Q = \frac{4\pi}{3}$ and after $Q = \frac{2\pi}{3}$, symmetric around $Q = + \pi$. The CSG boundary theory previously studied allowed no soliton emission or absorption, since the boundary could only have zero charge.

However the restriction to the theory with no field derivatives in the boundary term of the Lagrangian from the dressed boundary theory is by setting $A=0$ which actually amounts to setting $\alpha' = \alpha^{A=0} = \left\{ 0, \ \pm \pi \right\}$. This restricted theory is different to the original one studied with the conserved charge of this theory having the form
\begin{eqnarray}\label{eq:chargezero}
Q^{A=0} &=& i\int^{0}_{-\infty} dx \ \ \frac{u\partial_{t}u^{*} -u^{*}\partial_{t}u}{1-uu^{*}}  +\left.\left[2\alpha^{A=0}  \right]\right|_{x=0}\, ,
\end{eqnarray}
which allows the boundary to have charge $Q^{A=0} = \left\{ 0, \ \pm 2\pi \right\}$, not just zero charge. This theory does permit soliton emission and absorption, with the maximally charged soliton emitted from the excited defect and absorbed by the unexcited defect.

\subsubsection{Dressed boundary bound states}

Following on from analysing soliton absorption and emission, we will go on a ask whether there exists any dressed boundary bound states. Mathematically constructed by solving the boundary conditions with a stationary soliton solution in the bulk, the boundary conditions are satisfied when 
\begin{eqnarray}\label{eq:boundconstraint}
\left( \delta+\frac{1}{\delta}\right) &=& 2\frac{ \mathrm{cos}(a) \mathrm{sinh}(C) \sqrt{ \mathrm{cos}^{2}(A) \mathrm{cosh}(C)^{2} - \mathrm{cos}^{2}(a)} + \mathrm{sin}(a)\mathrm{sin}(A)\mathrm{cosh}(C)^{2}}{\mathrm{cosh}(C)^{2} -\mathrm{cos}^{2}(a)}\, ,
\end{eqnarray} 
where $C = 2\sqrt{\beta}\mathrm{cos}(a) c$. This constraint only makes sense with $\delta + \frac{1}{\delta}$ in the range \begin{equation} -\cos(A+a) < \frac{1}{2}\left(\delta + \frac{1}{\delta}\right) < \cos(A-a)\, ,\end{equation} which implies that $\delta$ is a pure phase. The energy and charge of the dressed bound state are calculated by substituting the stationary soliton solution into the total energy (\ref{eq:boundaryenergy}) and charge (\ref{eq:boundarycharge}) respectively. Simplifying using the constraint (\ref{eq:boundconstraint}) one finds
\begin{eqnarray}
E &=&  4\sqrt{\beta}|\mathrm{cos}(a)| + 4\sqrt{\beta}\mathrm{sin}(A)\mathrm{sin}(b)\, ,
\end{eqnarray}
 where $ \frac{1}{2}(\delta+\frac{1}{\delta}) = \mathrm{cos}(b)$ and
\begin{eqnarray}
Q &=& \pi -2a - 2\mathrm{\arcsin}\left(\frac{1}{2}(\delta+\frac{1}{\delta})\right)\, .
\end{eqnarray} 
 
These dressed boundary bound states include the bound states found in the original boundary theory, with all the above quantities with $A=0$ agreeing to the previous quantities. The way the charge is written above suggests that it does not change in the $A=0$ limit, this is not the case due to the $A$ dependence in (\ref{eq:boundconstraint}). 
 
\subsubsection{Soliton reflection}

The classical time-delay for a soliton reflecting from the dressed boundary can be calculated using a two-soliton solution satisfying the boundary conditions. By specifying that one of the constituent solitons is left moving and the other right moving, the idea is that the right moving soliton represents the soliton before it reflects from the boundary and the left moving one, the soliton after reflection. To make sure that energy and charge conservation are satisfied the parameter choices $a_{1}= a_{2} = a$ and $\theta_{1} = -\theta_{2} = \theta$ are made. 

Substituting the expression for a two-soliton solution (\ref{eq:2sol}), along with the constituent one-soliton solutions (\ref{eq:1sol}) $u_{1}, \ u_{2}$ with $N_{1}=1, \ N_{2}= e^{if}, \ c_{1}=c,\ c_{2}=d$ ($N_{1}$ can be set to one, since it is only the phase difference between the two solitons that affects the scattering properties) into the boundary condition creates a long and complicated expression. Each element of the boundary condition can be expanded in a power series in exponentials of time. As the boundary conditions should hold for all time the individual terms can be used to allow constraint conditions to be obtained. Solving for $(\delta+\delta^{-1})$ and using the fact that this is real gives two constraints, the first by specifying the imaginary part of the whole expression vanishes
\begin{eqnarray}
\label{eq:imagcons}
i\left(e^{\theta}\mathrm{sin}(\zeta+A)+ e^{-\theta}\mathrm{sin}(\zeta-A) \right) &=& \frac{\mathrm{sinh}(\theta)}{\mathrm{cos}(a)}\left(e^{-ia}\mathrm{sinh}(\lambda-iA) - e^{ia}\mathrm{sinh}(\lambda+iA)\right)\, ,
\end{eqnarray}
 and the remaining real equation
\begin{eqnarray}
\label{eq:realcons}
\delta + \frac{1}{\delta} &=& \frac{\mathrm{sin}(a)\left(e^{\theta}\mathrm{sin}(\zeta+A) -e^{-\theta}\mathrm{sin}(\zeta-A) \right) -\mathrm{cosh}(\theta)\left(e^{ia}\mathrm{sinh}(\lambda+iA) + e^{-ia}\mathrm{sinh}(\lambda-iA)   \right)}{\mathrm{cosh}(\lambda)+\mathrm{cos}(\zeta)}\, , \nonumber\\
&&
\end{eqnarray}
 where $$e^{\lambda}=e^{2\sqrt{\beta}\mathrm{cos}(a)\mathrm{cosh}(\theta)(c+d)},\ \ \ \ \ e^{i\zeta}=e^{2i\sqrt{\beta}\mathrm{sin}(a)\mathrm{sinh}(\theta)(c+d)}e^{-i\phi}\, .$$ Expanding the constraint equations and solving for $\mathrm{cosh}(\lambda),\mathrm{sinh}(\lambda), \mathrm{cos}(\zeta), \mathrm{sin}(\zeta)$ quadratics in $\mathrm{cosh}(\chi)$ can be found in terms of either $\lambda$ or $\zeta$ by eliminating the other. These quadratics are as follows
\begin{eqnarray}
\label{eq:lambda}
0 &=& \mathrm{cosh}^{2}(\chi) +\frac{\mathrm{cosh}(\theta)}{\mathrm{sinh}(\lambda)} \left(e^{ia}\mathrm{cosh}(\lambda+iA) + e^{-ia}\mathrm{cosh}(\lambda-iA)\right)\mathrm{cosh}(\chi) \nonumber\\
&&+ \frac{2\mathrm{sinh}(\lambda)\mathrm{cosh}(\theta) + e^{2ia}\mathrm{sinh}(\lambda+2iA) + e^{-2ia}\mathrm{sinh}(\lambda-2iA)}{4\ \mathrm{sinh}(\lambda)} \, ,\nonumber\\
&& \\
\label{eq:zeta}
0 &=& \mathrm{cosh}^{2}(\chi) + \frac{\mathrm{sin}(a)}{\mathrm{sin}(\zeta)}\left(e^{\theta} \mathrm{cos}(\zeta+A) - e^{-\theta}\mathrm{cos}(\zeta-A) \right)\mathrm{cosh}(\chi)\nonumber\\
&& - \frac{\left(2\ \mathrm{cos}(2a)\mathrm{sin}(\zeta) + e^{2\theta}\mathrm{sin}(\zeta+2A) + e^{-2\theta}\mathrm{sin}(\zeta-2A)\right)}{4\ \mathrm{sin}(\zeta)} \, .
\end{eqnarray}
These quadratics can then be solved for $ e^{2\lambda}$ and $e^{2i\zeta}$ respectively and re-expressed, using $$e^{\lambda} = e^{2\sqrt{\beta}\mathrm{cos}(a)\mathrm{sinh}(\theta)\Delta t}\, , \ \ \ \ \ e^{i\zeta} = e^{-i\phi}e^{2i\sqrt{\beta}\mathrm{sin}(a)\mathrm{cosh}(\theta)\Delta t}\, ,$$as the time-delay
\begin{eqnarray}\label{eq:dbtimedelay}
\Delta t &=& \frac{1}{ 2\sqrt{\beta}\mathrm{cos}(a)\mathrm{sinh}(\theta)}\mathrm{ln} \left| \frac{\mathrm{sinh}\left(\frac{\theta-\chi}{2} + i\frac{a-A}{2}\right)\mathrm{sinh}\left(\frac{\theta+\chi}{2} + i\frac{a-A}{2}\right)}{\mathrm{cosh}\left(\frac{\theta-\chi}{2} + i\frac{a+A}{2}\right)\mathrm{cosh}\left(\frac{\theta+\chi}{2} + i\frac{a+A}{2}\right)}\right| \, ,\nonumber\\
&&
\end{eqnarray}
and phase factor
\begin{eqnarray}\label{eq:dbphase}
e^{i\phi} &=& -\left(\frac{\delta e^{i\alpha}+e^{\theta}e^{ia}}{e^{\theta}e^{ia}e^{i\alpha}-\delta}\right)\left(\frac{ e^{i\alpha}+\delta e^{ia} e^{\theta}}{1 - \delta e^{\theta}e^{i\alpha}e^{ia}}\right)e^{2\sqrt{\beta}(\mathrm{sinh}(\theta+ia)\Delta t} \, .\nonumber \\
&&
\end{eqnarray}   

\begin{figure}[!h]
\begin{center}
\includegraphics[width=0.75\textwidth]{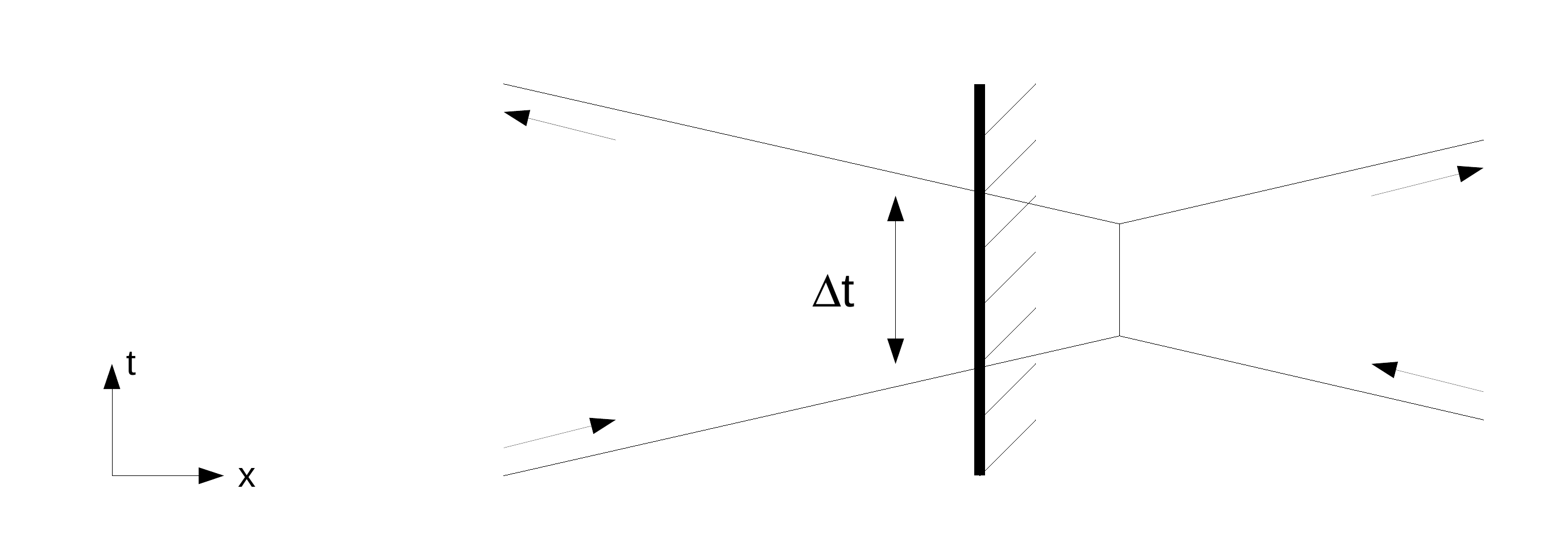}
\end{center}
\caption{Soliton reflection from dressed the boundary.}
\label{Fig:solitonreflection}
\end{figure}
As shown in figure \ref{Fig:solitonreflection} the time-delay of the reflected soliton is due to two effects, the standard time-delay from the soliton-soliton scattering and the position behind the boundary where this scattering takes place. These two effects both arise because of the mathematical construction used to describe the soliton scattering from the boundary, in reality the time-delay exhibited by the soliton is due to a non-zero time scattering with the boundary.

\subsubsection{Particle reflection}

To complete the analysis of solutions interacting with the boundary we will investigate how the CSG particle interacts with the dressed CSG boundary. To calculate the particle reflection matrix, we linearise the dressed boundary conditions giving
\begin{eqnarray}
\frac{\partial u_{\epsilon}}{\partial x} - i\ \mathrm{tan}(A) \frac{\partial u_{\epsilon}}{\partial t} - \frac{ \sqrt{\beta}}{\mathrm{cos}(A)} \left(\delta+ \frac{1}{\delta}\right) u_{\epsilon} &=&0\, ,\nonumber\\
\frac{\partial u^{*}_{\epsilon}}{\partial x} + i\ \mathrm{tan}(A) \frac{\partial u^{*}_{\epsilon}}{\partial t} - \frac{ \sqrt{\beta}}{\mathrm{cos}(A)} \left(\delta+ \frac{1}{\delta}\right) u^{*}_{\epsilon} &=&0\, .
\end{eqnarray}
 These linearised equations are then solved with an incoming and outgoing particle \begin{equation}u_{\epsilon} = u_{particle}(\theta) + R_{particle} u_{particle}(-\theta),\end{equation} which gives the particle reflection factor
\begin{eqnarray}
R_{particle} &=& \frac{2i\ \mathrm{sinh}(\theta+iA) + \delta+\frac{1}{\delta}}{2i\ \mathrm{sinh}(\theta-iA) - (\delta+\frac{1}{\delta})}\, .
\end{eqnarray}

\begin{figure}[h]
\begin{center}
\includegraphics[width=0.8\textwidth]{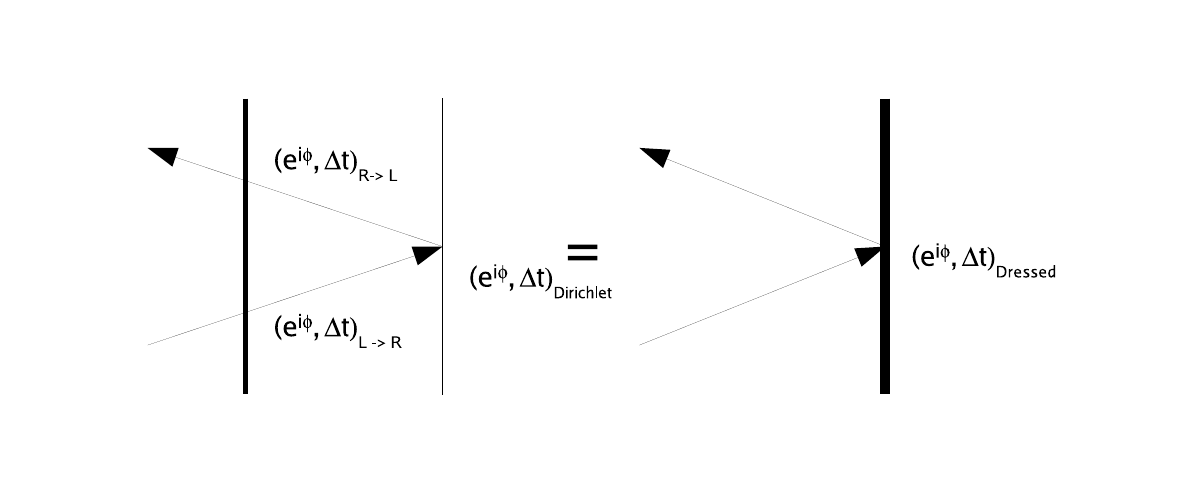}
\end{center}
\caption{Consistency check for soliton or particle reflection from the dressed boundary.}
\label{Fig:reflectioncheck}
\end{figure}

This particle reflection factor can also be obtained as the $a = \frac{\pi}{2}$ limit of the soliton reflection phase factor. By using the construction of the dressed boundary there is a consistency check, shown in figure \ref{Fig:reflectioncheck}, for the defect-soliton time-delay and dressed boundary-soliton time-delay and also soliton phase-factors and particle reflection factors. The diagram shows that the dressed boundary time-delay should be equal to the sum of the left to right defect time-delay, the Dirichlet boundary time-delay and the right to left defect time-delay \begin{equation} \Delta t_{dbound} = \Delta t _{defect}^{l \rightarrow r} +\Delta_{Dirichlet} +\Delta t _{defect}^{l \rightarrow r} \, ,\end{equation} the quantities derived in this paper do satisfy this relation. The soliton and particle phase factors satisfy similar relations.

\section{Summary and Discussion}
In this paper we have shown how defects can be introduced for the complex sine-Gordon model and how the defect can be used to create a new CSG boundary theory. This was facilitated by finding an explicit formula for $\alpha$ (\ref{eq:alpha}), which dispenses with the need to consider the fields and dual fields simultaneously. This enabled both the B\"{a}cklund transformation and two-soliton solution to be written in terms of only the fields, with the BT now having an extra parameter $A$.

By construction the defect conditions at $x=0$  are identical to the BT, following the pattern of defects in previously studied integrable theories. As a check on the integrability of these conditions, explicit formulae for the conserved energy, momentum and charge in the presence of a defect were constructed. Like the sine-Gordon theory the CSG defect can be excited or unexcited which enables the defect to absorb and emit solitons when the soliton and defect parameters are particularly related. In SG these processes are possible because the defect can store the discrete topological charge, whereas in CSG the defect can store the continuous $U(1)$ Noether charge. In both theories there are two different defects that can absorb the same soliton, one defect with $\delta >0$ and the other with $\delta<0$. In SG these two defects have different topological charges and the CSG defects different charge $Q$.

The CSG soliton can scatter through the defect experiencing a time-delay (or time-advance) along with a phase shift and the particle scatters through the defect with no reflection. This purely transmitting nature of the defect is so far a universal property of integrable defects.   

The integrable defect was used to construct a dressed boundary theory which also maintains classical integrability. The dressed boundary provides a wider class of  boundary conditions to those previously studied, which can be recovered as a specific limit of the new dressed boundary. As expected the dressed boundary inherits properties from its constituent defect, namely the absorption and emission of particular solitons. A classical bound state is presented along with the soliton and particle reflection factors.

\begin{figure}[h]
\begin{center}
\includegraphics[width=0.6\textwidth]{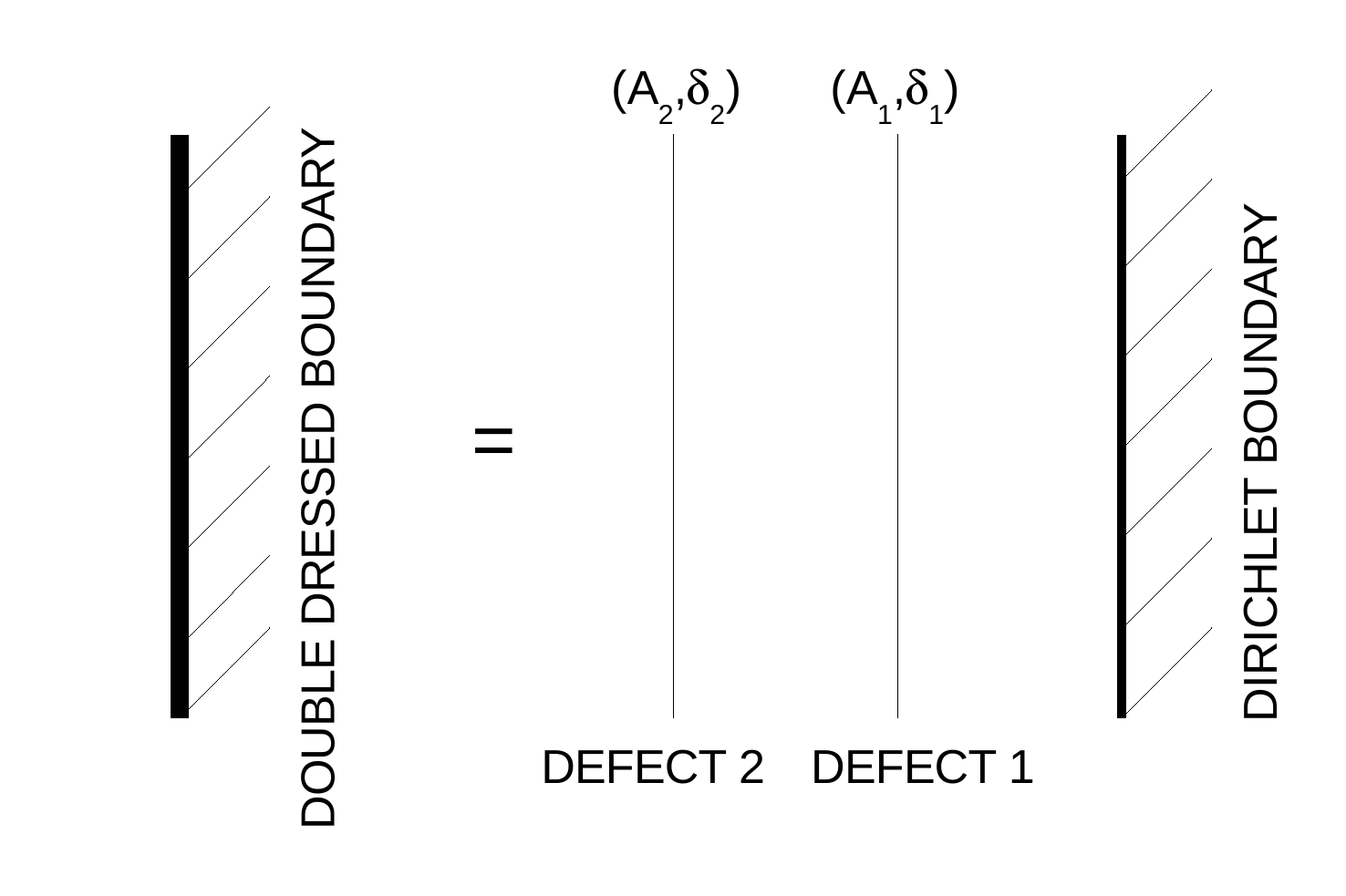}
\end{center}
\caption{The construction of a double dressed boundary.}
\label{Fig:doubledressed}
\end{figure}

Repeating the construction of the dressed boundary, another boundary could be produced by putting two defects in front of the Dirichlet boundary. To explicitly write down the boundary conditions of such a doubly dressed boundary is computationally tricky and beyond the scope of this work. However the properties of such a boundary are expected to include the ability to absorb solitons when the parameters match either of the defects. It is interesting to ponder whether two defects with the same parameters are allowed to placed alongside each other unlike the exclusion principle which does not allow either identical SG or CSG solitons to be placed alongside one another.

Placing defects in front of the boundary act like a soliton filter; it is possible to create an arbitrary absorption spectrum by placing appropriate defects in front of the boundary. In fact, one might speculate 
that there is the hypothetical limit where by placing an infinite number of different defects in front of the boundary one could create a boundary that was fully absorbing.

The work in this paper thoroughly analyses the classical aspects of both the CSG model with defect and dressed boundary. Future work will concentrate on the quantum versions of both these theories.  Also, as mentioned in the introduction, integrable CSG boundaries  have been used in a string theory context \cite{Ahn:2007bq} and it is an interesting question to ask whether integrable defects might also play a role in string and gauge theories?

\subsubsection{Acknowledgements}
We would like to thank Ed Corrigan, Patrick Dorey and Paul Sutcliffe for discussions. 
JMU thanks the Engineering and Physical Sciences Research Council for a PhD studentship.

\bibliographystyle{hunsrt}
\bibliography{CSGDefects}

\end{document}